\documentclass{aa}
\usepackage{txfonts}
\usepackage{amsmath}
\usepackage{color}
\usepackage{xcolor}
\usepackage[normalem]{ulem}
\usepackage{multicol}
\usepackage[colorlinks=true,citecolor=blue,linkcolor=blue]{hyperref}
\usepackage{textcomp}
\usepackage{subcaption}
\usepackage{graphicx}
\usepackage{float}
\usepackage{placeins}
\usepackage{float}
\usepackage{multirow}
\usepackage{changepage}
\usepackage{colortbl}

\author{
	J.M.G.H.J. de Jong\inst{\ref{inst:Leiden}} 
	\and H.J.A. R\"ottgering\inst{\ref{inst:Leiden}}
	\and R. Kondapally\inst{\ref{inst:edin}}
	\and B. Mingo\inst{\ref{inst:open}}
	\and R.J. van Weeren\inst{\ref{inst:Leiden}} 
	\and P.N. Best\inst{\ref{inst:edin}}
	\and L.K. Morabito\inst{\ref{inst:durham},\ref{inst:durham2}}
	\and M. Magliocchetti\inst{\ref{inst:rome}}
	\and J.B.R. Oonk\inst{\ref{inst:surf},\ref{inst:Leiden},\ref{inst:Astron}}
	\and A. Villarrubia-Aguilar\inst{\ref{inst:Leiden}}
	\and F. F. Vecchi\inst{\ref{inst:Leiden},\ref{inst:eth}}}

\institute{
	Leiden Observatory, Leiden University, PO Box 9513, 2300 RA Leiden, The Netherlands\relax\label{inst:Leiden} \and 
	Institute for Astronomy, University of Edinburgh, Royal Observatory, Blackford Hill, Edinburgh, EH9 3HJ, UK \relax\label{inst:edin} \and
	SURF/SURFsara, Science Park 140, 1098 XG Amsterdam, The Netherlands\relax\label{inst:surf} \and 
	School of Physical Sciences, The Open University, Walton Hall, Milton Keynes MK7 6AA, UK\relax\label{inst:open} \and
	ASTRON, The Netherlands Institute for Radio Astronomy, Postbus 2, 7990 AA Dwingeloo, The Netherlands\relax\label{inst:Astron} \and
	ETH Zurich, Institute for Particle Physics and Astrophysics, Wolfgang-Pauli-Str 27, Zurich 8093, Switzerland\relax\label{inst:eth} \and
	INAF-IAPS, Via Fosso del Cavaliere 100, 00133, Rome, Italy\relax\label{inst:rome} \and
	Centre for Extragalactic Astronomy, Department of Physics, Durham University, Durham DH1 3LE, UK\relax\label{inst:durham} \and
	Institute for Computational Cosmology, Department of Physics, University of Durham, South Road, Durham DH1 3LE, UK\relax\label{inst:durham2}
}
\begin{document}
	
	\title{Cosmic evolution of FRI and FRII sources out to $z=2.5$}
	
	\date{Received XXX 2023 / Accepted XXX 2023}
	
	\begin{abstract} 
		{
			Radio-loud active galactic nuclei (RLAGN) play an important role in the evolution of galaxies through the effects on their environment. The two major morphological classes are core-bright (FRI) and edge-bright (FRII) sources. With the LOw-Frequency ARray (LOFAR) we can now compare the FRI and FRII evolution down to lower flux densities and with larger samples than before. 
		}
		{
			Our aim is to examine the cosmic space density evolution for FRIs and FRIIs by analyzing their space density evolution between $L_{150}\sim 10^{24.5}$~W~Hz$^{-1}$ and $L_{150}\sim 10^{28.5}$~W~Hz$^{-1}$ and up to $z=2.5$. We look in particular at the space density enhancements and compare the FRI and FRII evolution with the total RLAGN evolution.
		}
		{
			We construct radio luminosity functions (RLFs) from FRI and FRII catalogues based on recent data from LOFAR at 150~MHz to study the space densities as a function of radio luminosity and redshift. These catalogues contain over 100 times the number of FRIs with associated redshifts greater than $z=0.3$ compared to similar FRI/FRII RLF studies.
			To derive the maximum distance a source can be classified and to correct for detection limits, we simulate how sources appear at a range of redshifts.
		}
		{
			Our RLFs do not show any sharp transitions between the space density evolution of FRI and FRII sources as a function of radio luminosity and redshift. We report a space density enhancement from low to high redshift for FRI and FRII sources brighter than $L_{150}\sim 10^{27}$~W~Hz$^{-1}$. Furthermore, while we observe a tentative decrease in the space densities of FRIs with luminosities below $L_{150}\sim 10^{26}$~W~Hz$^{-1}$ and at redshifts beyond $z=0.8$, this may be due to residual selection biases. The FRI/FRII space density ratio does not appear to evolve strongly as a function of radio luminosity and redshift.
		}
		{
			We argue that the measured space density enhancements above $L_{150}\sim 10^{27}$~W~Hz$^{-1}$ are related to the higher gas availability in the earlier denser universe. The constant FRI/FRII space density ratio evolution as a function of radio luminosity and redshift suggests that the jet-disruption of FRIs might be primarily caused by events occurring on scales within the host galaxy, rather than being driven by changes in the overall large-scale environment. The remaining selection biases in our results also highlight the need to resolve more sources at angular scales below 40\arcsec and therefore strengthens the motivation for the further development and automation of the calibration and imaging pipeline of LOFAR data to produce images at sub-arcsecond resolution.
		}
		\maketitle
	\end{abstract}
	
	\section{Introduction}
	
	
	It is believed that most, if not all, galaxies are hosting a super massive black hole (SMBH) \citep{magorrian1998, kormendy2013}, where a part of those are powered by an accretion disk around it \citep{lyndell1969, kauffmann2003, martini2013}. We classify these objects as active galactic nuclei (AGN). They are called radio-loud AGN (RLAGN) when they also produce powerful collimated jets emitting at radio frequencies, due to gas falling onto the SMBH which interacts with its magnetic field, generating synchrotron emission \citep[see review by][and references therein]{hardcastle2020}. RLAGN play an important role in the evolution of the universe, as their jets heat their environment, accelerate high energy cosmic rays, and might be responsible for a big part of the intergalactic magnetic field \citep{blandford2019}. In addition, RLAGN influence the evolution of their host galaxy through feedback processes, where the energy released by the AGN can prevent cooling or expel gas. This can regulate or even quench star formation and therefore affect the growth of the host galaxies \citep[e.g.,][]{croton2006, best2006, mcnamara2007, cattaneo2009, fabian2012, morganti2017, hardcastle2020}. Moreover, there also seems to be a link between the radio loudness and the host morphology, where the most powerful RLAGN are hosted by massive geometrically roundly shaped galaxies \citep{barisic2019, zheng2020, zheng2022}. All of the above are reasons to argue for the importance to study the cosmic evolution of RLAGN.
	
	RLAGN morphologies can be classified based on the distance from the central galaxy to the brightest point of their jets \citep{fr}. The two main classes are the core-bright FRI morphologies and the edge-bright FRII morphologies. \citeauthor{fr} found in their sample a `break luminosity' at radio power $L_{150}\sim 10^{26}$~W~Hz$^{-1}$, where FRIs dominate below this luminosity value and FRIIs above. In the study by \cite{ledlow}, it was demonstrated that the break luminosity increases for increasing host optical luminosity. This observation suggested a direct link between the morphology of radio jets and their environmental density. As a result, it was proposed that RLAGN initially exhibit an FRII morphology but transition to an FRI morphology when they are unable to traverse the local interstellar medium, which breaks the jet-collimation and decelerates the jet speed on kpc-scales from their host galaxy \citep{bicknell1995, kaiser2007}. More recent studies, based on samples of radio sources less affected by selection effects compared to older studies, showed that the break luminosity does not separate FRI and FRII morphologies as strictly as was initially observed \citep{best2009, gendre2010, gendre2013, wing2011, capetti2017b, mingo2019, mingo2022}. The FRI/FRII morphological divide is therefore less closely connected to radio luminosities than previously suspected.
	Nonetheless, the existence of FRII below the original break luminosity remains consistent with the jet-disruption model if this population is a mix of restarting FRIIs, (old) fading FRIIs, or FRIIs hosted by less massive host galaxies in less dense environments compared to FRIIs above the break luminosity \citep{croston2019, mingo2019}. Examining the cosmic evolution of FRI and FRII morphologies therefore links to studying the evolution of the large-scale environments of radio galaxies \citep[e.g.][]{croston2019}.
	
	The changing interpretation of the break luminosity demonstrates how selection biases play an important role in the ability to detect FRI and FRII sources and as a result affect our understanding of radio galaxy evolution. Because FRIIs have hotspots at the edge of the lobes and are more powerful, they will be easier to detect than FRIs. This selection effect becomes stronger when we look at more distant RLAGN, where their jets become fainter and are closer to the flux density limit from the used instrument of the survey. It has been debated how much of the observed evolution of the FRI and FRII sources is due to selection effects and how much is due to true evolution of the radio galaxy population \citep{singal2014, magliocchetti2022}. The difficulty in detecting FRIs at high redshifts, with very limited sample sizes, led to the prediction that powerful FRIs should be significantly more abundant at $z>1$ compared to what we find locally \citep{snellen2001, jamrozy2004, rigby2008}. A larger study of the FR dichotomy, with the combined NVSS-FIRST (CoNFIG) catalogue \citep{gendre2008}, strengthened this prediction by finding positive space density enhancements from low to high redshift up to a factor of 10 over the local population for both FRIs and FRIIs up to $z=2.5$ \citep{gendre2010, gendre2013}. To understand the role of extended RLAGN across cosmic time, it is vital to fully characterise the cosmic evolution of the FRI and FRII sources using much larger samples and down to fainter radio luminosities than previous studies.
	
	In this paper, we investigate the FRI/FRII space density evolution as a function of redshift and radio luminosity by constructing radio luminosity functions (RLFs) up to $z=2.5$ with the catalogues from \cite{mingo2019, mingo2022}. These catalogues are based on the 150-MHz LOw-Frequency ARray \citep[LOFAR;][]{haarlem2013} surveys that  have a high surface brightness sensitivity and are therefore great at detecting FRIs with low flux densities. By correcting more precisely for selection effects through redshift simulations, we derive better estimates of the maximum volume that a source can still be observed and classified over, before falling below the detection and selection limits of the sample. These simulations in combination with the theoretical angular size distribution, help us to better correct for the incompleteness of over sample such that we can recover an improved estimate of the `real' FRI/FRII RLFs. This comprehensive approach along with 100 times more FRI redshifts above $z=0.3$ compared to prior work from \cite{gendre2010}, also aid to improve the FRI and FRII RLF comparisons above this redshift and to re-examine the findings from \citeauthor{gendre2010} regarding the space density enhancements of FRIs and FRIIs up to $z=2.5$.
 
	In Section \ref{sec:data} of this paper we briefly discuss the selected data and catalogues. This is followed in section \ref{sec:redshiftsimulation} by an explanation of our redshift simulation that we use in section \ref{sec:rlf} to construct RLFs from which we derive our results in section \ref{sec:results}. We discuss our results in section \ref{sec:discussion} and look at future prospects to further improve this work. Finally, we conclude in section \ref{sec:conclusion}. 
	For our work, we use a $\Lambda$CDM cosmology model with $H_0=70$~km~s$^{-1}$~Mpc$^{-1}$, $\Omega_{m}=0.3$, and $\Omega_{\Lambda}=0.7$.
	
	\section{Data}\label{sec:data}
	
	We utilize in this paper FRI/FRII sources classified by \citet[][their catalogue is hereafter referred to as M19]{mingo2019} combined with a similarly compiled catalogue based on deeper observations from \citet[][their catalogue is hereafter referred to as M22]{mingo2022}. Both catalogues are constructed with data from the LOFAR Two-metre Sky Survey \citep[LoTSS][]{shimwell2017, shimwell2019} at 144~MHz with a resolution of 6\arcsec and a sensitivity ranging from \textasciitilde 20 to \textasciitilde 70 $\mu$Jy\,beam$^{-1}$. The sensitive low-frequency observations used to construct these catalogues excel at identifying complex extended sources, which benefits the detection of extended diffuse jets from FRIs. We will discuss in the following subsections their content and our source selection.
	
	\subsection{Catalogues}
	M19 contains sources from LoTSS DR1, based on radio maps from the HETDEX Spring Field covering 424 deg$^{2}$ with a median sensitivity of 71 $\mu$Jy\,beam$^{-1}$ \citep{shimwell2017, shimwell2019, williams2019}, while M22 is containing sources from the LoTSS-Deep Fields DR1 which are about 3 to 4 times deeper than LoTSS DR1 and covers 25\,deg$^2$ \citep{tasse2021, sabater2021, kondapally2021, duncan2021, best2023}. The deep fields were selected with deep wide-area multi-wavelength imaging from ultraviolet to far-infrared (see \cite{kondapally2021} for details) and are consisting of the following three fields:
	European Large Area Infrared Space Observatory Survey-North 1 (ELAIS-N1) \citep{oliver2000}, Lockman Hole \citep{lockman1986}, and Bo\"otes \citep{jannuzi1999}.
	The corresponding LoTSS-Deep Fields DR1 radio maps have individually a median sensitivity of about 20, 22, and 32 $\mu$Jy beam$^{-1}$ for ELAIS-N1, Lockman Hole, and Bo\"otes respectively \citep{tasse2021, sabater2021}.
	
	The M19 and M22 catalogues contain for every source their total 150-MHz flux density, size, and host galaxy redshift and position on the sky. The sizes and radio flux densities from sources in M19 and M22 were carefully measured by adopting a flood-filling procedure and by comparing their sizes and flux densities with what was found with the \texttt{PyBDSF} Gaussian fitting tool \citep[See for more details Section 2.5 in][]{mingo2019}.\footnote{\url{https://pybdsf.readthedocs.io/}} These comparisons were followed up by visual inspection. The sources from the catalogue from LoTSS DR1 to construct M19 are for 73 \% identified with an optical host \citep{williams2019} and have 51 \% spectroscopic or photometric redshifts \citep{duncan2019}. Sources in the deep fields were over \textasciitilde 97\% identified with radio host-galaxies associated with carefully determined photometric redshifts (or spectroscopic where available) using a hybrid approach of template fitting and machine learning methods \citep{duncan2019, duncan2021}. \citeauthor{mingo2022} included in M22 only sources from the deep fields up to $z=2.5$, as the spectral energy distribution (SED) fitting of all the detected radio sources in the deep fields was considered reliable for sources up to this redshift \citep{best2023}. The FRI and FRII classifications of these sources were done using the \texttt{LoMorph} classification code and by additional visual inspection \citep{mingo2019}.\footnote{\url{https://github.com/bmingo/LoMorph}} They reported a classification accuracy of 89\% for FRIs and 96\% for FRIIs. \citeauthor{mingo2019} excludes sources with projected angular sizes smaller than 27\arcsec~or below 40\arcsec~with a 20\arcsec or smaller distance between their two peak emissions. This angular size cut is based on classification difficulties and the resolution limits from LoTSS DR1 and the LoTSS-Deep Fields DR1.
	Also other sources that did not clearly fit in FRI or FRII sources were filtered out \citep[see Section 2.4 from][]{mingo2019}. Those contain double-double sources (restarting FRII) \citep{schoenmakers2000} and hybrid sources \citep{gopalkrishna2000}.
	
	\subsection{Source selection}\label{sec:dataselection}
	The sources from M19 with reliable redshifts above $z=0.8$ are primarily quasars. This is mainly due to the scarcity of spectroscopic data for radio galaxies and the line-of-sight bias \citep{duncan2019, hardcastle2019}. As a result, we have excluded sources with redshifts beyond $z=0.8$ from M19. We do not need to make additional redshift cuts for M22, as the radio galaxies from M22 are drawn from the significantly deeper LoTSS-Deep Fields DR1, in contrast to LoTSS DR1, which have precise redshifts up to $z=2.5$ \citep{kondapally2021, duncan2021, best2023}. Consequently, all the sources utilized in this paper with redshifts between $z=0.8$ and $z=2.5$ originate exclusively from M22 and are only \textasciitilde 9\% of our full sample. Although the sources collected from the LoTSS-Deep Fields DR1 have a higher redshift completeness than the sources collected from LoTSS DR1, we find the ratio of the number of FRI/FRIIs for the overlapping redshifts below $z<0.8$ for M19 and M22 to be similar. Also, the redshift distributions below $z<0.8$ are similar, i.e. we find their redshifts to have the same mean and standard deviation. This demonstrates that combining M19 and M22 does not introduce additional selection effects as a function of redshift.
	
\begin{figure*}[ht]
		\centering
		\includegraphics[width=1\linewidth]{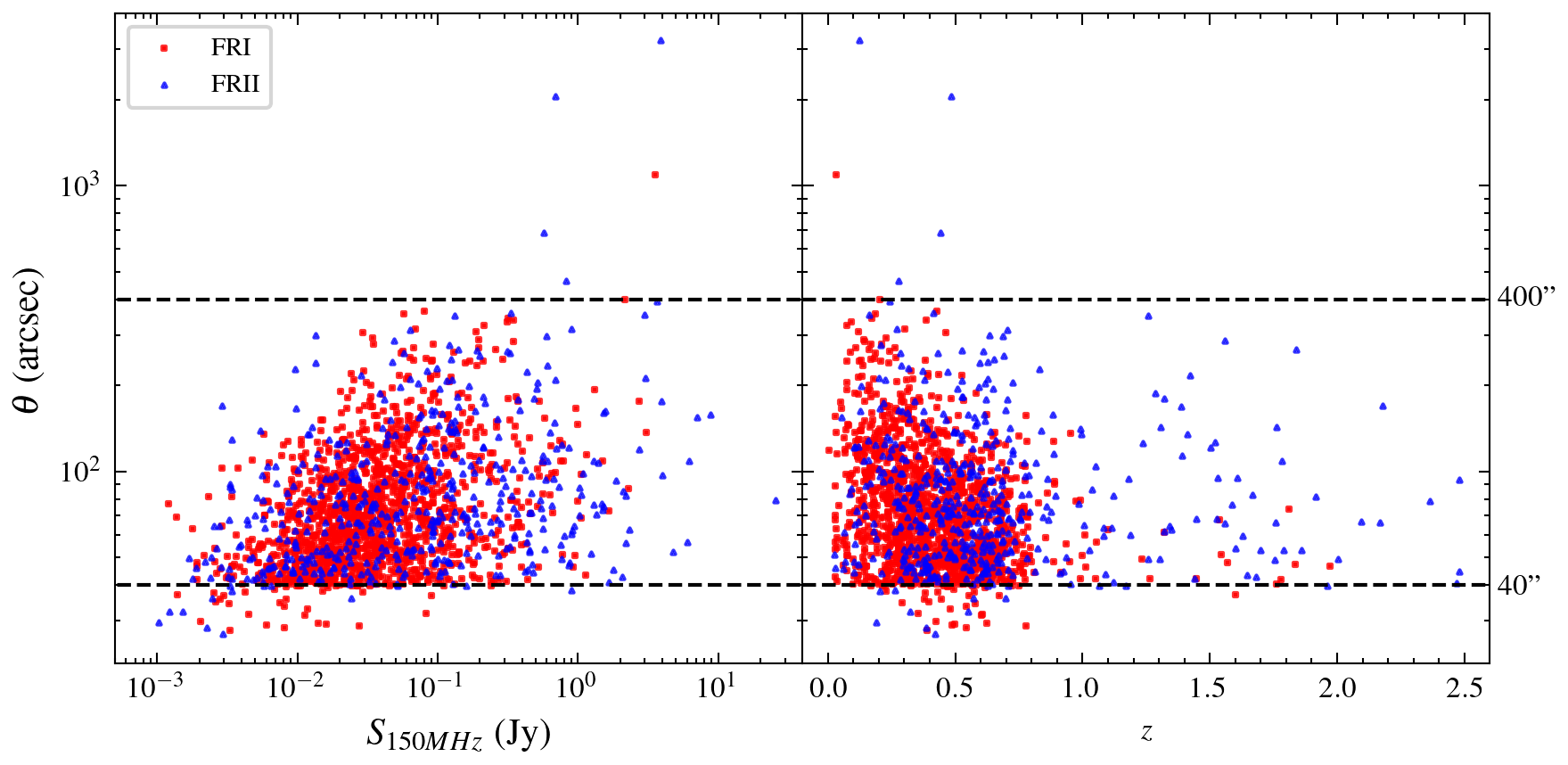}
		\caption{Angular size as a function of flux density (left panel) and redshift (right panel) for the 1560 sources considered in this paper. We separated our sources into 1146 FRI and 414 FRII morphologies with separated colors and draw dashed lines for $40\arcsec$ and $400\arcsec$. These lines determine the angular size cuts that we applied in this paper before doing completeness corrections (see Section \ref{sec:completeness}).}
		\label{fig:angulsizedistr}
	\end{figure*}

	In Figure \ref{fig:angulsizedistr} we plot the angular size versus the radio flux density and redshift for all our FRIs and FRIIs. From this figure we see that, due to the selection criteria by \citeauthor{mingo2019}, there are very few sources between 27\arcsec~and 40\arcsec in the sources from M19 and M22. Most of those sources are below $z=0.8$. We also find a clear gap of sources above 400\arcsec, where only 0.3\% of our sources are situated. This is the area where classification of bright giant radio galaxies (GRGs) becomes difficult because different selection biases for FRIs and FRIIs are playing a role in relation to the surface brightness sensitivity of the used observation. 
	On the one hand, are the most distant parts of the FRI jets closer to or below the surface brightness limit, which in some cases makes only their unresolved core appear or their sizes being measured smaller than they actually are. This issue becomes more prominent at the higher end of our redshift distribution where surface brightness limits play a more dominant role due to the fact that sources are on average fainter as a function of redshift. This effect can also be observed in the right panel of Figure \ref{fig:angulsizedistr}, where FRIs are closer positioned to the 40\arcsec~boundary compared to FRIIs at higher redshifts. On the other hand, might large FRIIs with jets close to the surface brightness limit appear to only have two bright disconnected hotspots. This poses challenges in associating them with each other, particularly when their separation is substantial.
	So, it is not unexpected that a significant proportion of large sources are missing in our catalogues \citep[e.g.][]{oei2023}. To account for the 40\arcsec~and 400\arcsec~angular size limits, we will exclude sources that fall below and above these threshold values, and subsequently implement completeness corrections as a function of angular size and flux density to regain the contribution of sources beyond these boundaries to our RLFs, as we will discuss in more detail in section \ref{sec:angularsizecorrections}. 
	
\begin{figure}[ht]
		\centering
		\includegraphics[width=1\linewidth]{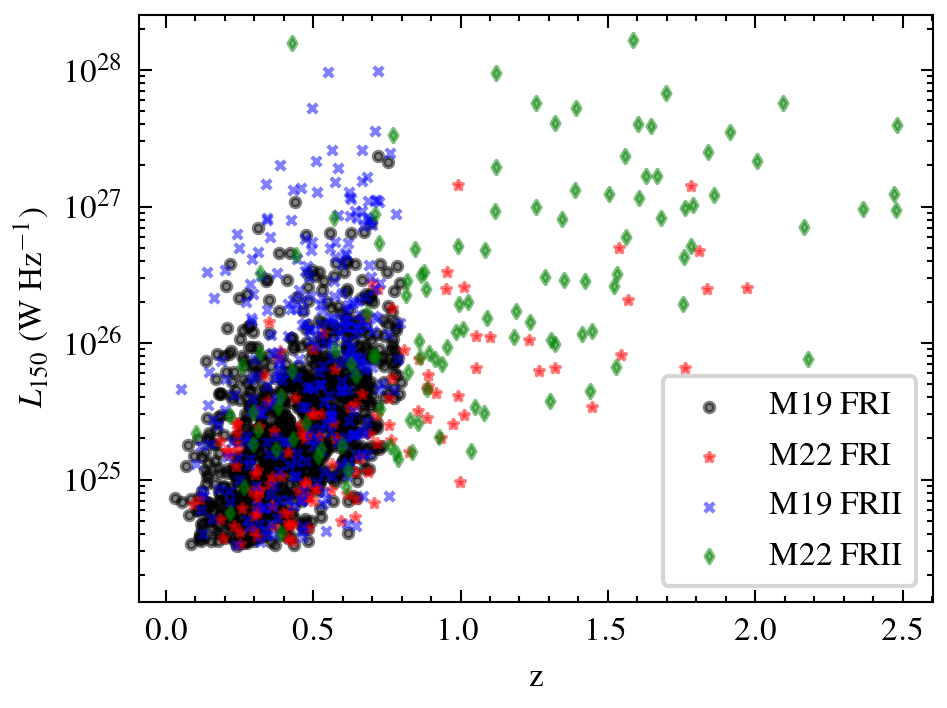}
		\centering
		\caption{Luminosity-redshift diagram for M19 and M22 split in FRIs and FRIIs.}
		\label{fig:lumzplot}
	\end{figure}

	We further limit our analyses to sources with a radio luminosity above $L_{150}\sim 10^{24.5}$~W~Hz$^{-1}$, as the number of sources below this radio luminosity is small. This also ensures that our completeness corrections are not strongly affected by star-forming galaxies \citep[e.g.][]{sabater2019, cochrane2023} or compact sources \citep[e.g.][]{sadler2016, baldi2018, compact}. Moreover, due to the LoTSS flux density limits, we do not find beyond $z=0.8$ any sources below $L_{150}\sim 10^{24.5}$~W~Hz$^{-1}$. So, this decision improves the reliability of completeness corrections and therefore our resulting RLFs without compromising the goal of this paper, which is to investigate the evolution of radio galaxies across redshifts up to $z=2.5$. Figure \ref{fig:lumzplot} shows the final selection of the 1560 sources with their 150~MHz radio luminosities as a function of redshift. From those are 1146 classified as FRI sources and 414 as FRII sources.

	\section{Redshift simulations}\label{sec:redshiftsimulation}
	
	Detection limits from radio telescopes, restrict our view of the universe. In order to incorporate these effects into the derivation of RLFs for FRIs and FRIIs, we have developed an algorithm that simulates how sources would appear after relocating them to higher redshifts, such that we can derive up to which maximum distance a source remains reliably classifiable. These redshift simulations will also help us to determine the incompleteness of our sample as a function of flux density and redshift.
	
	\subsection{Surface brightness and redshift relation}
	
	We can relate the total flux density ($S_{\nu}$) of a source to its luminosity ($L_{\nu}$) and its luminosity distance ($d_{L}$) by
	\begin{equation}\label{eq:fluxdens}
		S_{\nu}=\frac{L_{\nu}(1+z)^{1-\alpha}}{4\pi d_{L}^{2}},
	\end{equation}
	where $(1+z)^{1-\alpha}$ accounts for the K-correction with spectral index $\alpha$ and where the radio flux density is $S_{\nu}\propto \nu^{-\alpha}$.
	The angular size $\theta_{z}$ of a source is related to its physical size $s$ and the angular distance ($d_{A}$) by
	\begin{eqnarray}\label{eq:angulardist}
		\theta_{z} & = & \frac{s}{d_{A}}  \nonumber \\
		& = & \frac{s}{d_{L}}\cdot(1+z)^{2},
	\end{eqnarray}
	where we used the relation $d_{A}=\frac{d_{L}}{(1+z)^{2}}$.
	Combining the above, we derive the surface brightness evolution over redshift by
	\begin{eqnarray}\label{eq:sb}
		\Sigma & = & \frac{S_{\nu}}{\theta_{z}^{2}} \nonumber \\
		& = & \frac{L_{\nu}}{4\pi s^{2}}\cdot (1+z)^{-3-\alpha}.
	\end{eqnarray}
	As the luminosity and physical size are physical intrinsic properties of a radio galaxy, this demonstrates that the surface brightness decreases over redshift by a factor $(1+z)^{3+\alpha}$.
	
	\subsection{Redshifting algorithm}\label{sec:redshifting}
	
	If we could relocate a source to a higher redshift, we would see the angular size and flux density from the source in the image change. The source would diminish in size according to the adopted cosmology until approximately $z \sim 1.6$, after which it would begin to increase once again. These angular size changes affect the ability to detect and classify FRIs and FRIIs at the same radio powers differently, due to their different morphologies. On the one hand, the FRI will be classifiable until the diffuse jets are not bright enough to be detected or the source cannot be resolved anymore due to its small angular size. On the other hand, the FRII remains classifiable as long as the hotspots appear as point sources and the components can be associated with the host, you have the resolution to separate them, and they are still bright enough above the noise. Additionally, there are also situations where FR classifications can change after redshift increments from FRI to FRII and vice versa, due to the detection limits and the changing number of pixels covering the source.
	To simulate and better understand these effects, we developed the \texttt{redshifting} algorithm to increment their redshift by $\Delta z$.\footnote{\url{https://github.com/jurjen93/redshifting}} 
	
	The \texttt{redshifting} algorithm takes as input an image of a source at its observed redshift. After every redshift increment, the algorithm changes the pixel sizes by multiplying the pixel scale by $\frac{\theta_{z}}{\theta_{z+\Delta z}}$, which is equivalent to dividing the pixel scale by the ratio between $d_{A}$ at $z$ and $d_{A}$ at $z+\Delta z$. This information is used to resample the image pixels with the \cite{reproject} resampling algorithm.\footnote{\url{https://github.com/astropy/reproject/}} 
	In this way, the source appears smaller or larger after redshift increments, depending on the redshift.
	Following from Equation \ref{eq:sb}, the algorithm reduces every pixel value by a factor $\left(\frac{1+z}{1+z+\Delta z}\right)^{3+\alpha}$. Because the image noise is independent from the source's redshift, we keep it constant during our procedure of relocating a source to a higher redshift. This means for our algorithm that we only reduce pixel values that exceed our $3\sigma$ noise threshold. 
	
	We adopt a constant spectral index of $\alpha=0.7$ for the K-correction. This spectral index value is the typical average spectral index for RLAGN and often used to construct radio luminosity functions if the spectral indices of sources are unknown \citep[e.g.][]{condon2002, mauch2007, padovani2016, prescott2016, hardcastle2016, calistro2017, kondapally2022}. A typical spectral index variation between $\alpha=0.6$ or $\alpha=0.8$ \citep[e.g.][]{hardcastle2016, calistro2017, murphy2017} would, compared to $\alpha=0.7$, only adjust the pixel brightness changes from our redshifting algorithm up to \textasciitilde~5\% for redshift increments of $\Delta z=0.7$. This is the mean $\Delta z$ from their original $z$ we find sources still to be classifiable after applying the redshifting algorithm (see following sections) for the sources considered in this paper. Hence, it is not expected that the choice for $\alpha=0.7$ will strongly bias our results. 
 
    We also make sure that the total power remains constant as a function of redshift increments. This implies that we do not correct for inverse Compton (IC) scattering losses from the cosmic microwave background (CMB), which is expected to add additional selection biases \citep{krolik1991, morabito2018, sweijen2022b}. We will further discuss in Section \ref{sec:interpretingrlfs} the implications of this choice on the interpretation of our RLFs when we do not consider this effect.
	
	\begin{figure*}[ht]
		\begin{subfigure}{.33\textwidth}
			\includegraphics[width=1\linewidth]{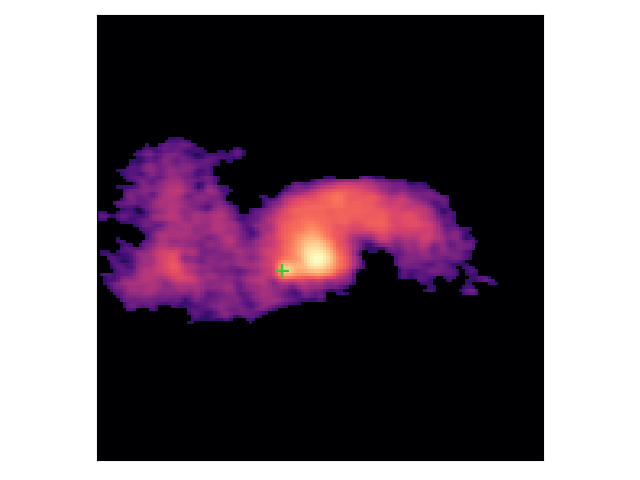}
			\caption{FRI at $z=0.48$}
		\end{subfigure}
		\begin{subfigure}{.33\textwidth}
			\includegraphics[width=1\linewidth]{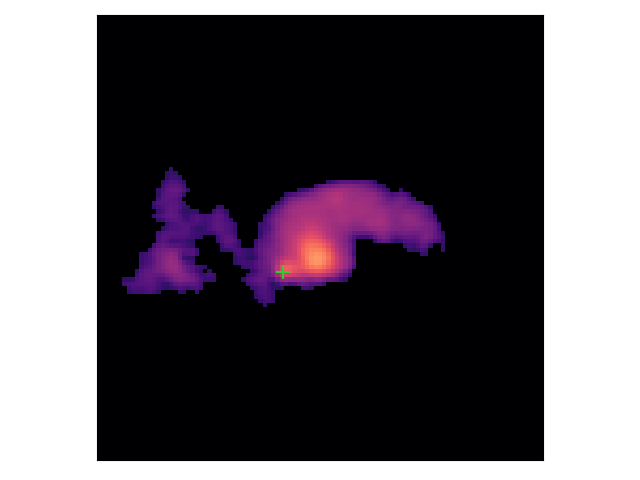}
			\caption{FRI at $z=0.93$}
		\end{subfigure}
		\begin{subfigure}{.33\textwidth}
			\includegraphics[width=1\linewidth]{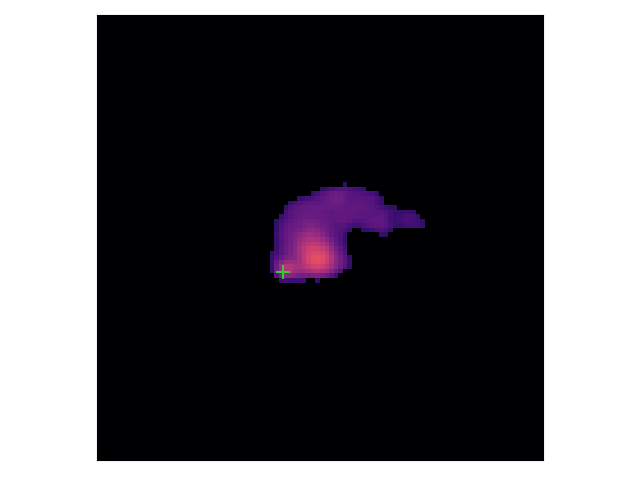}
			\caption{FRI at $z=1.38$}
		\end{subfigure}
		\begin{subfigure}{.33\textwidth}
			\includegraphics[width=1\linewidth]{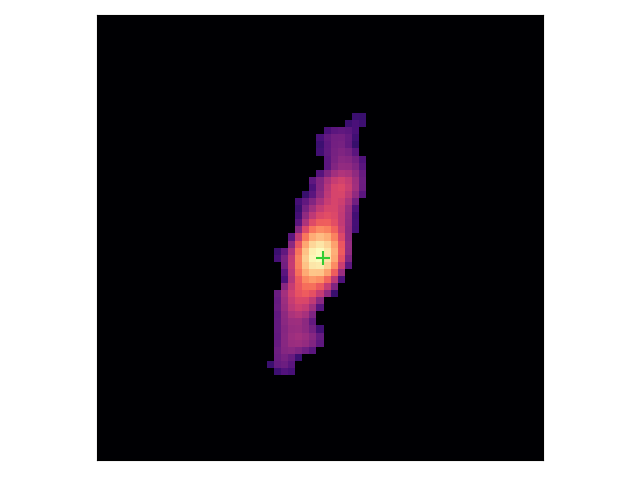}
			\caption{FRI at $z=0.19$}
		\end{subfigure}
		\begin{subfigure}{.33\textwidth}
			\includegraphics[width=1\linewidth]{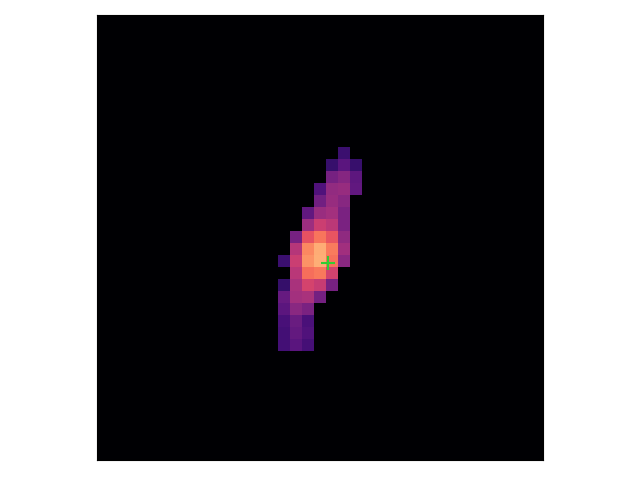}
			\caption{FRI at $z=0.39$}
		\end{subfigure}
		\begin{subfigure}{.33\textwidth}
			\includegraphics[width=1\linewidth]{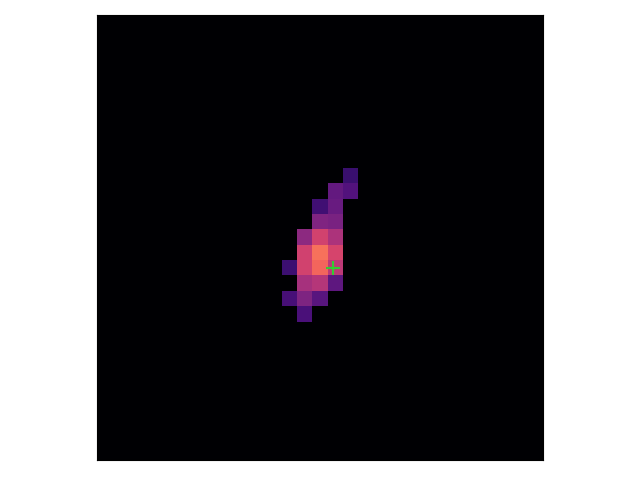}
			\caption{FRI at $z=0.59$}
		\end{subfigure}
		\begin{subfigure}{.33\textwidth}
			\includegraphics[width=1\linewidth]{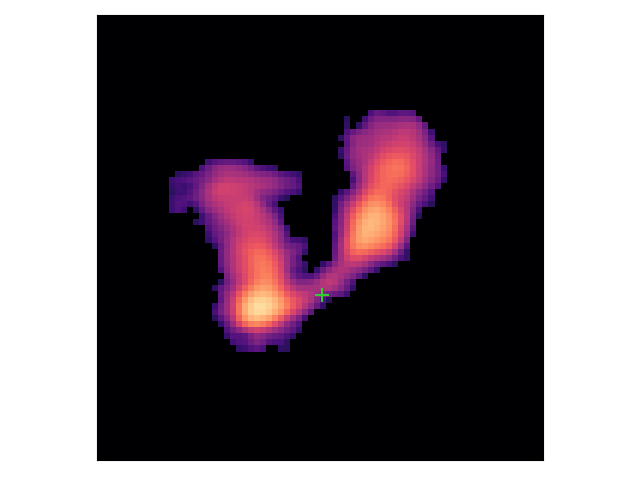}
			\caption{FRI at $z=0.29$}
		\end{subfigure}
		\begin{subfigure}{.33\textwidth}
			\includegraphics[width=1\linewidth]{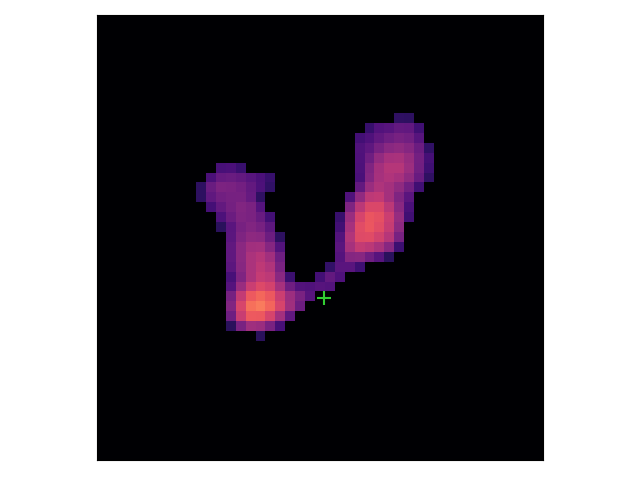}
			\caption{FRI at $z=0.69$}
		\end{subfigure}
		\begin{subfigure}{.33\textwidth}
			\includegraphics[width=1\linewidth]{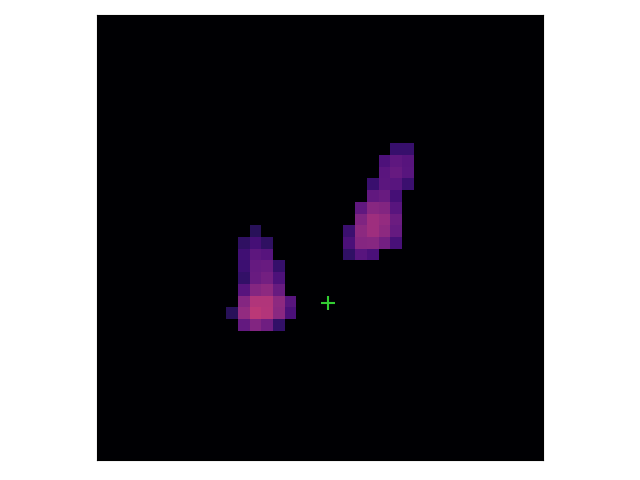}
			\caption{FRI at $z=1.19$}
		\end{subfigure}
		\caption{Examples of FRI sources after redshift increments by applying the \texttt{redshifting} algorithm. All unassociated emission is masked out (in this case below $4\sigma$). Each columns displays the source from the same row at different redshifts. The last row is a wide-angle tail FRI source \citep{owen1976}, where after redshift increments the source appears to look like an FRII source. The optical host location is in every image given by the green cross.}
		\label{fig:redshift_examples_fri}
	\end{figure*}
	
	\begin{figure*}[ht]
		\begin{subfigure}{.33\textwidth}
			\includegraphics[width=1\linewidth]{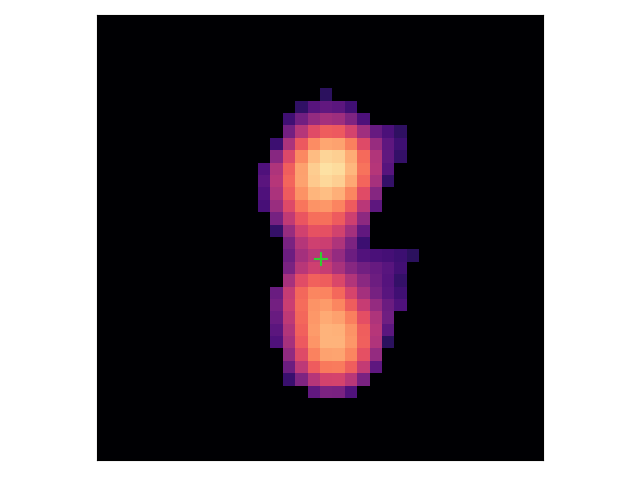}
			\caption{FRII at $z=0.33$}
		\end{subfigure}
		\begin{subfigure}{.33\textwidth}
			\includegraphics[width=1\linewidth]{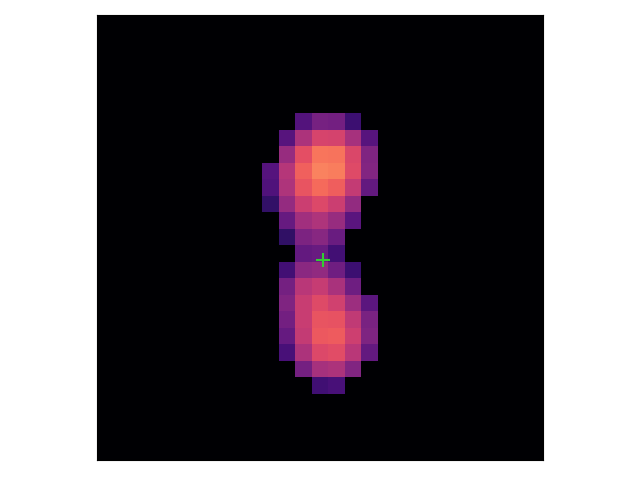}
			\caption{FRII at $z=0.83$}
		\end{subfigure}
		\begin{subfigure}{.33\textwidth}
			\includegraphics[width=1\linewidth]{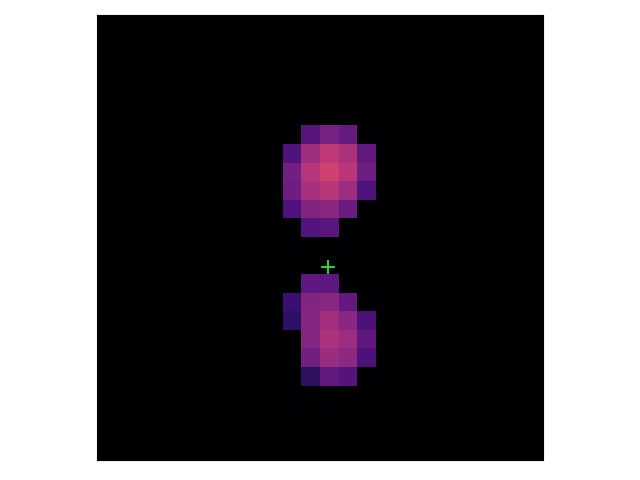}
			\caption{FRII at $z=1.33$}
		\end{subfigure}
		\begin{subfigure}{.33\textwidth}
			\includegraphics[width=1\linewidth]{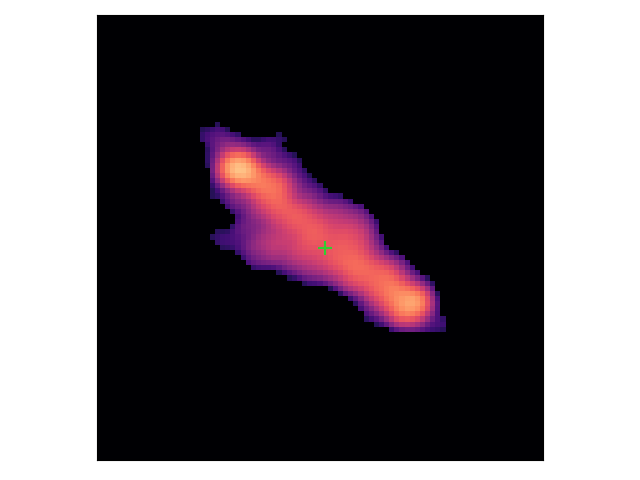}
			\caption{FRII at $z=0.9$}
		\end{subfigure}
		\begin{subfigure}{.33\textwidth}
			\includegraphics[width=1\linewidth]{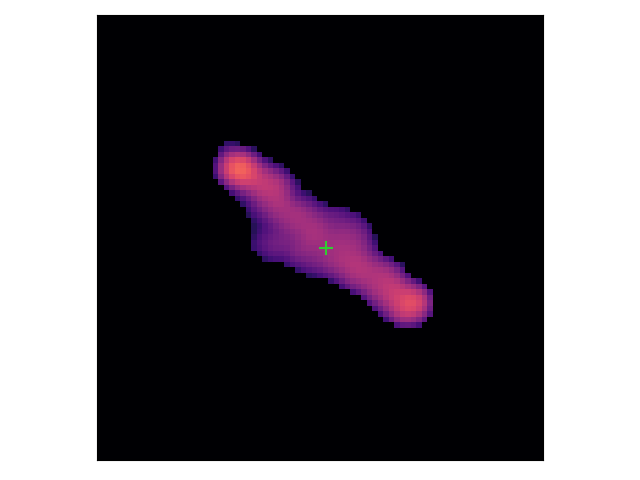}
			\caption{FRII at $z=1.5$}
		\end{subfigure}
		\begin{subfigure}{.33\textwidth}
			\includegraphics[width=1\linewidth]{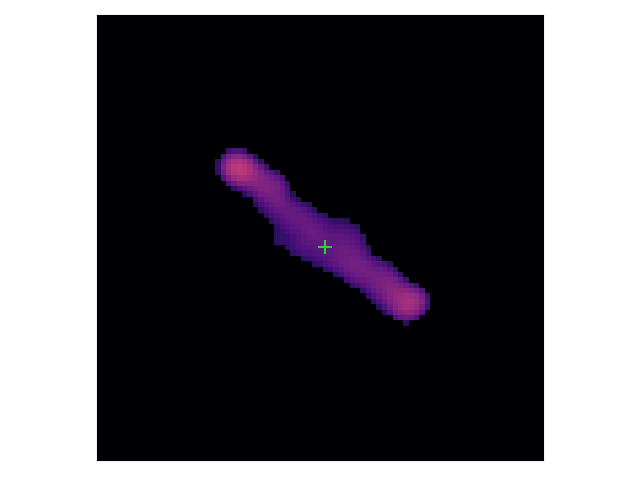}
			\caption{FRII at $z=2.1$}
		\end{subfigure}
		\begin{subfigure}{.33\textwidth}
			\includegraphics[width=1\linewidth]{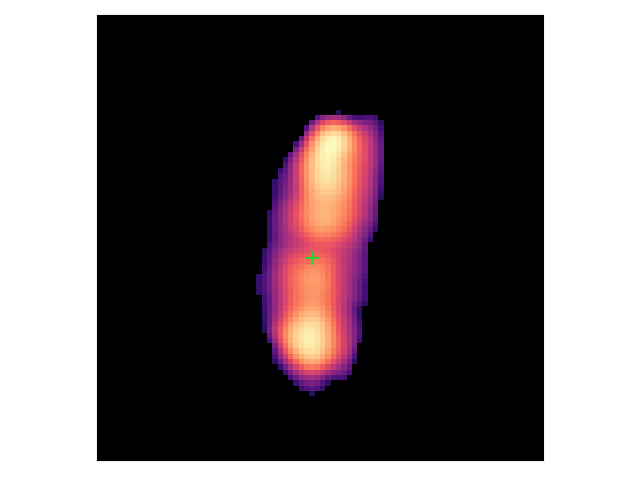}
			\caption{FRII at $z=0.34$}
		\end{subfigure}
		\begin{subfigure}{.33\textwidth}
			\includegraphics[width=1\linewidth]{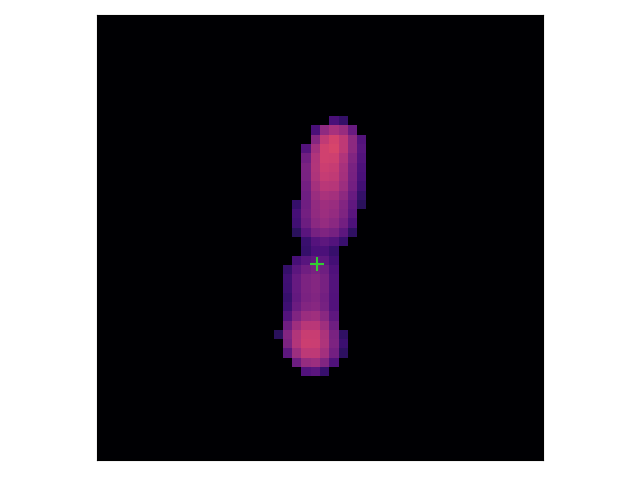}
			\caption{FRII at $z=1.34$}
		\end{subfigure}
		\begin{subfigure}{.33\textwidth}
			\includegraphics[width=1\linewidth]{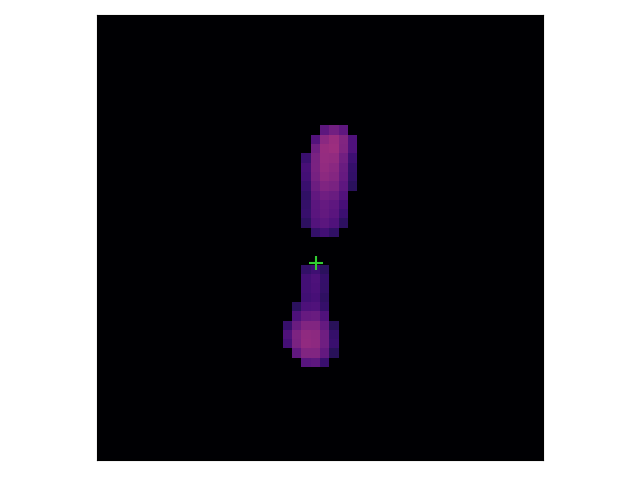}
			\caption{FRII at $z=1.84$}
		\end{subfigure}
		\caption{Examples of FRII sources after redshift increments by applying the \texttt{redshifting} algorithm. All unassociated emission is masked out (in this case below $4\sigma$). Each columns displays the source from the same row at different redshifts. The optical host location is in every image given by the green cross.}
		\label{fig:redshift_examples_frii}
	\end{figure*}
	
	\subsection{Source components}
	
	The association of radio source components is a difficult task and massive visual inspection through citizen science projects or neural network architectures are often used to do this for large amounts of data \citep{banfield2015, williams2019, mostert2022}.\footnote{See also the LOFAR Galaxy Zoo project: \url{https://www.zooniverse.org/projects/chrismrp/radio-galaxy-zoo-lofar}} For all our sources we already have initial information about the source components \citep[based on \texttt{PyBDSF} and visual inspection, see][]{mingo2019}. So, an additional step for associating source components is not necessary, as our focus is solely on determining the components that remain above the noise threshold at each redshift increment. All unassociated components in the image are masked out in every source cutout image. This is necessary when multiple radio sources are close to each other, or when there are remaining islands of emission contaminating the image for classification.
	After relocating a source to a new redshift with \texttt{redshifting}, we draw polygons around the remaining components with emission above a particular noise level.
	
	The sources in Figures \ref{fig:redshift_examples_fri} and \ref{fig:redshift_examples_frii} demonstrate the effect of the redshifting increments on the observed morphology for different FRIs and FRIIs. We see in particular how their brightness reduce and how their appearance change. The first two rows of Figure \ref{fig:redshift_examples_fri} display FRI sources where their jets partly disappear below the $\sigma$-threshold when they are relocated at higher redshifts. In the last row we have a special case where an FRI source appears as an FRII after large redshift increments.
	The sources in Figure \ref{fig:redshift_examples_frii} are typical cases of FRIIs with clear hotspots. These sources are showing how the FRII hotspots are the main components that remain visible after relocating them at higher redshifts. We can see from the first and the last row how the elongated FRII component separates into two components at higher redshifts.
	
	\subsection{FR classification}\label{sec:classification}
	
	Because we use only sources from M19 and M22, we based our FRI/FRII classification code on the \texttt{LoMorph} code, which were used to classify our sources. This classification algorithm is based on measuring the distance from the optical host to the brightest region and the edge of the source by using flood-filling and verifying if a source is resolved or unresolved by applying \texttt{PyBDSF}. For a more detailed discussion we refer to Section 2 from \cite{mingo2019}.
	
	The classification of the source on the third row in Figure \ref{fig:redshift_examples_fri} changes both visually and in our algorithm from an FRI to an FRII morphology after being incremented to a higher redshift. This classification inconsistency occurs for about \textasciitilde 2\% of all our sources when relocating them to higher redshifts. FRIIs could also become classified as FRIs, when, due to the resampling of pixels, the hotspots of an FRII merge into a bright FRI type core. We did not find these cases in our simulations, which indicates that before this would occur, the source was already not classifiable anymore. This could also be because double-double and hybrid sources have already been all filtered out from our sources (see Section \ref{sec:dataselection} and \cite{mingo2019, mingo2022}).
	These classification inconsistencies are rare and prove that our morphological classification is rather effective.
	
	\section{Constructing radio luminosity functions}\label{sec:rlf}
	
	RLFs are an important tool to study the evolution of radio sources as a function of power and redshift. We follow the standard method from \cite{schmidt1968} for the construction of the RLFs in this paper. This means that we calculate the density evolution as a function of redshift $z$ and radio luminosity $L$ by
	\begin{equation}\label{func:rlf}
		\rho(z, L) = \frac{1}{\Delta \log L}\sum_{n=1}^{N} \frac{1}{V_{\text{max,n}}},
	\end{equation}
	where $V_{\text{max,n}}$ is the maximum volume that source $n$ can be classified and where $\Delta \log L$ is the log value of the radio luminosity bin. In the next subsections we further discuss the $V_{\text{max}}$ and the completeness corrections we apply to construct the RLFs.
	
	\subsection{$V_{\text{max}}$ method}
	
	To determine $V_{\text{max}}$ we first need to find an accurate value for the maximum redshift a source can still be classified ($z_{\text{max}}$). This value can be found by applying the \texttt{redshifting} algorithm we discussed in Section \ref{sec:redshifting}.
	
	\subsubsection{Determining $z_{\text{max}}$}\label{sec:zmax}
	
	We find $z_{\text{max}}$ by spatially relocating each source across the radio image 200 times and re-evaluating up to which maximum redshift the source is still classifiable at each spatial location with the \texttt{redshifting} algorithm ($\tilde{z}_{\text{max}}$). In this way, we get a reliable measure for $z_{\text{max}}$ by taking the mean of all the obtained $\tilde{z}_{\text{max}}$ values and can find the error on this value by taking the standard deviation of the $\tilde{z}_{\text{max}}$ values ($\sigma_{z_{\text{max}}}$). The value for $\sigma_{z_{\text{max}}}$ embeds information about the effects of the varying noise in the radio maps. 
	
\begin{figure}[ht]
		\centering
		\includegraphics[width=1\linewidth]{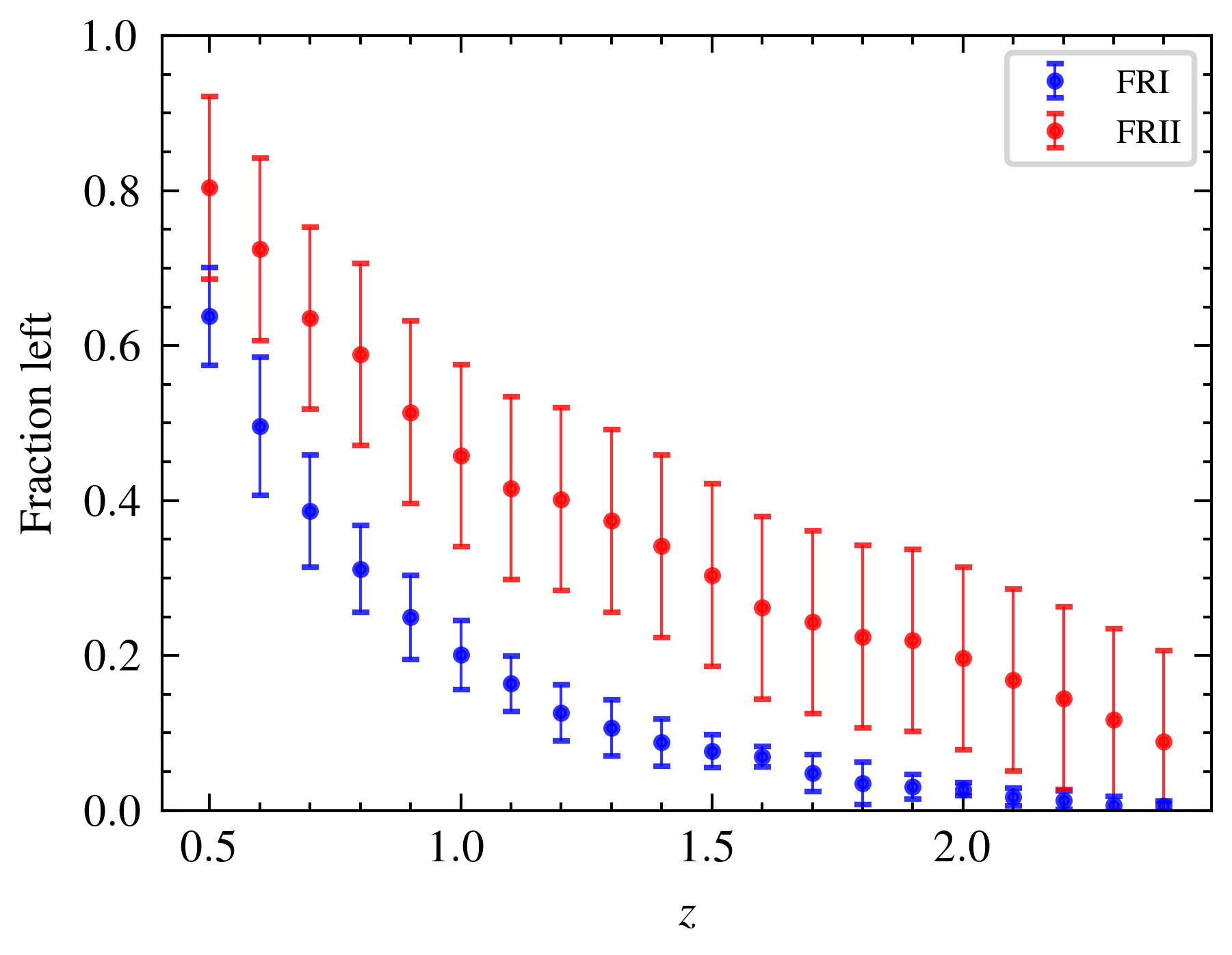}
		\caption{Fraction of local ($z<0.5$) FRI and FRII sources that are still classifiable after applying redshift increments with the \texttt{redshifting} algorithm as a function of redshift.  The error bars include the errors on $z_{\text{max}}$ and Poisson errors.}
		\label{fig:FRIFRII_redshift}
	\end{figure}

\begin{figure}[h]
		\centering
		\includegraphics[width=1\linewidth]{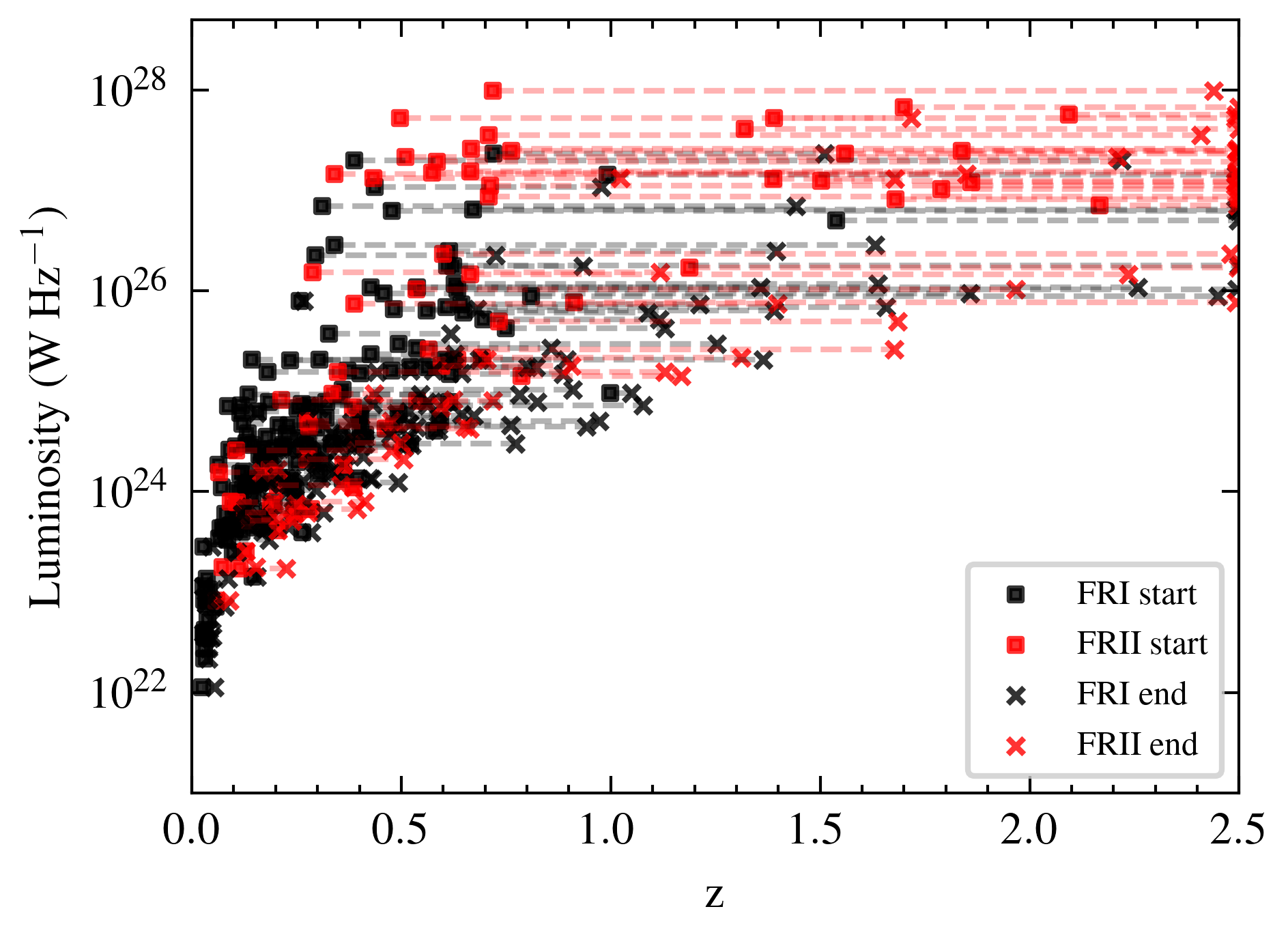}
		\caption{Luminosity versus redshift increments with the \texttt{redshifting} algorithm for 150 random sources at different luminosities from all sources from M19 and M22 above $L_{150}\sim 10^{22}$~W~Hz$^{-1}$. The starting positions are at the original measured redshifts given by the squared boxes, while the ending positions are the final maximum redshifts a source can be classified (up to $z=2.5$). }
		\label{fig:lum_z}
	\end{figure}

	With the obtained $z_{\text{max}}$ values, we can compare how FRIs and FRIIs are differently affected by the \texttt{redshifting} algorithm. For this, we can for example take a local sample up to $z=0.5$ and move those up to $z=2.5$. In Figure \ref{fig:FRIFRII_redshift} we show from this example the fraction of the remaining sources that are still detectable and classifiable in steps of $z=0.1$. Only 50\% of the FRI sources we start with remain classifiable up to $z\sim0.6$, while for FRII sources we still classify 50\% of the sources up to $z\sim1$. This is partly explained by the different effects detection limits have on the observed morphologies of FRI and FRII sources (see Section \ref{sec:redshifting}). In addition, FRII sources are on average also brighter as they are locally more abundant above the break luminosity \citep{fr, mingo2019}. We demonstrate this in Figure \ref{fig:lum_z} where we show how 150 randomly picked sources move in $P-z$ space from their original observed $z$-position to their $z_{\text{max}}$ position. On the higher luminosity end, we observe, as expected, how FRII sources reach higher $z_{\text{max}}$ values.
	
	\subsubsection{Final integral for $V_{\text{max}}$}\label{sec:vmax}
	
	The $z_{\text{max}}$ determines the maximum redshift a source can be observed in for each redshift bin $z_{\text{bin}}$ over which we evaluate the following integral
	\begin{equation}\label{eq:vmax}
		V_{\text{max}} =\int_{z_{\text{bin}}} C(S_{\nu}) \Theta(z)\chi(z)V(z)\text{d}z,
	\end{equation}
	where $C(S_{\nu})$ is the completeness correction as a function of flux density2 at a particular redshift (see Section \ref{sec:completeness}), $V(z)\text{d}z$ is the infinitesimal comoving volume slice between $z$ and $z+dz$, $\Theta(z)$ is the fractional sky coverage, and $\chi(z)$ the binary function
	$$
	\chi(z) = \left\{
	\begin{array}{ll}
		1 & \mbox{if $z\leq z_{\text{max}}$} \\
		0 & \mbox{if $z> z_{\text{max}}$}.
	\end{array}
	\right.
	$$
	The uncertainties on $z_{\text{max}}$ propagate in $V_{\text{max}}$ by calculating $\tilde{V}_{\text{max}}$ corresponding to every $\tilde{z}_{\text{max}}$ with Equation \ref{eq:vmax} and taking the standard deviation of all the obtained values.
	The final uncertainties of the space densities also include Poisson errors following from \cite{gehrels1986} and the \texttt{LoMorph} classification accuracies from \cite{mingo2019} (89\% for the FRIs and 96\% for the FRIIs).
	
	As Figure \ref{fig:FRIFRII_redshift} demonstrates how the FRIs are more strongly affected by selection effects, this figure also shows how with our $V_{\text{max}}$ method the FRI space densities will be more strongly corrected as a function of $z$ compared to the space densities from FRII sources.
	
	\subsection{Completeness corrections}\label{sec:completeness}
	
	The $V_{\text{max}}$ method enables measuring the space densities for sources within our sample as a function of redshift. We also need to apply a correction that takes into account the sources that are undetected due to the flux density and resolution limits from our observations. This is the (in)completeness factor as defined in Equation \ref{eq:vmax} by $C(S_{\nu})$. For this work, this means that we need to estimate how many FRI and FRII sources we are missing as a function of flux density ($S_{150}$), such that we can better estimate the real space density values. Firstly, we calculate the completeness corrections by generating mock sources based on the sources from our source sample. Secondly, we determine the completeness correction for the sources with sizes below 40\arcsec~and above 400\arcsec. Combining both corrections gives our final completeness correction.
	
	\subsubsection{Correction for $40\arcsec<\theta<400\arcsec$}\label{sec:cor1}
	
	We find the completeness corrections from our sources as a function of flux density by first generating mock FRI and FRII sources.
	Those mock sources are created by multiplying the flux density from a source in our catalogue with a random factor loguniformly drawn between 0 and 1 and scaling the pixel brightness in our image with the same factor. We then replace the mock source to the same noise environments as the noise effects are already embedded in the $\sigma_{z_{\text{max}}}$ values. When we do this 200 times per source, we have about \textasciitilde 350k mock sources. Using the classification part from Section \ref{sec:classification}, we decide for each case if it is still recoverable as an FRI or FRII source. As the M19 and M22 samples have different sensitivities, we do this for both catalogues separately. With these simulations we find the flux density incompleteness for sources corresponding to the angular size distribution from the objects in our combined M19 and M22 catalogues.
	
	\subsubsection{Correction for $\theta<40\arcsec$ and $\theta>400\arcsec$}\label{sec:angularsizecorrections}
	
	As explained in Section \ref{sec:dataselection}, we miss sources smaller than 40\arcsec~and larger than 400\arcsec.
	To quantify the angular size incompleteness in our completeness weights, we use the empirical integral angular size distribution from \cite{windhorst1990} with the updated parameters fitted to LoTSS-Deep Fields DR1 by \cite{mandal2021}. This is a radio source angular size distribution which has been used in multiple studies and has proven to be a reliable measure \citep{prandoni2001, huynh2005, hales2014, mahony2016, williams2016, retana2018, prandoni2018, mandal2021, kondapally2022}. 
	The angular size distribution follows by
	\begin{equation}\label{eq:windhorst}
		\Psi(>\theta) = \text{exp}\left(-\ln{2}\left(\frac{\theta}{\Tilde{\theta}}\right)^{0.8}\right),
	\end{equation}
	where $\theta$ is in arcseconds and where $\Tilde{\theta}$ is the median angular size which \citeauthor{windhorst1990} derived to be related to the flux density $S_{1400}$ in mJy by
	\begin{equation}\label{eq:mediantheta}
		\Tilde{\theta} = 2\cdot k\cdot \left(S_{1400}\right)^{m},
	\end{equation}
	where $m=0.3+0.2\cdot\text{exp}\left(-S_{1400}^2\right)$ and 
	$$
	k = \left\{
	\begin{array}{ll}
		7-3\cdot \text{exp}\left(-\frac{S_{1400}}{2}\right) & \mbox{if $S_{1400} < 4.5$} \\
		4+3\cdot \text{exp}\left(-\frac{S_{1400}}{200}\right)  & \mbox{if $S_{1400}\geq 4.5$},
	\end{array}
	\right.
	$$
	where we doubled the value of $k$ compared to \cite{mandal2021}, as \cite{kondapally2022} showed for LoTSS-Deep Fields data the angular size distribution to fit well for AGN with twice the median angular size. By converting Equation \ref{eq:mediantheta} from $\nu=1.4$~GHz to $\nu=150$~MHz with spectral index $\alpha=0.7$ and combining this with Equation \ref{eq:windhorst}, we find the angular size distribution as a function of the flux density $S_{150}$.
	
	The radio luminosity cut below $L_{150} = 10^{24.5}$~W~Hz$^{-1}$ (see Section \ref{sec:dataselection}) ensures that the angular size completeness corrections are dominated by extended radio sources. Nonetheless, we cannot distinguish with the angular size distribution from \citeauthor{windhorst1990} the distributions of FRIs and FRIIs. Fortunately, considering the comparison of FRI and FRII physical size distributions by \cite{best2009}, it is reasonable to infer that both morphologies have comparable size distributions. \cite{best2009} state with data from FIRST and NVSS that above 40 kpc the size distributions of FRI and FRII radio galaxies are similar, whereas the smallest source in our sample is 83 kpc and 95\% of our sources are larger than 230 kpc. However, the measured angular sizes from FRIs are likely to be more underestimated at higher redshifts, as sources become dimmer as a function of redshift, which affects the detectibility of the full extend of diffuse FRI jets. As a result, the angular size completeness corrections for FRIs might become less accurate for higher redshifts, compared to FRIIs whose sizes are defined by their brightest components. However, to simplify the construction of the RLFs, we rely only on the measured angular sizes to determine the angular size completeness corrections and do therefore not derive a redshift dependency. The effect of this issue on our RLFs will be discussed in Section \ref{sec:totalagn}.
	
	\subsubsection{Final completeness correction}
	
\begin{figure}[ht]
		\centering
		\includegraphics[width=1\linewidth]{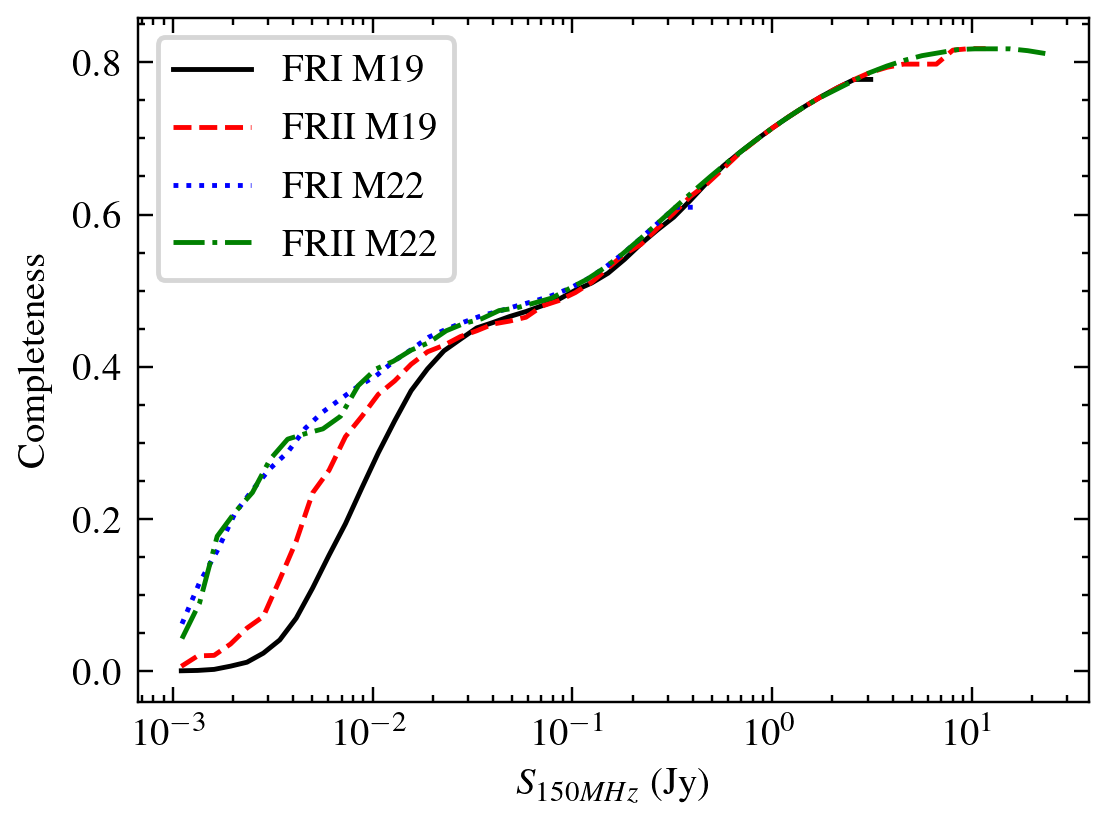}
		\centering
		\caption{Completeness corrections for FRI and FRII sources from M19 and M22.}
		\label{fig:complete}
	\end{figure}

	We present our final full completeness corrections (including all corrections discussed above) as a function of flux density in Figure \ref{fig:complete}. We identify a 50\% completeness around \textasciitilde 100 mJy. This is also after which the completeness corrections start to be dominated by the angular size completeness corrections. This becomes evident in the convergence of the completeness for FRI and FRII sources at the upper end of the flux densities, stemming from the assumption of comparable size distribution for both FRIs and FRIIs.
	The flattening at the end of the completeness curve is showing the effects of the 400\arcsec~upper angular size threshold that starts to dominate. After 10~Jy, the completeness corrections start to fall off, which matches with the assumption that a notable fraction bright GRGs are missing due to selection effects (see Section \ref{sec:dataselection}). Below \textasciitilde~100~mJy we notice how the better surface brightness sensitivity in LoTSS-Deep Fields DR1, compared to LoTSS DR1, improves the completeness from M22. For M19 we also notice how FRII sources at lower flux densities are more complete compared to FRI sources. This is because the FRI jets tend to be diffuser and more likely to disappear below the detection limit, while an FRII with the same flux density will typically remain classifiable as long as their prominent hotspots remain visible.
	Because the completeness corrections become extremely large at the lower radio flux density end, we set, similar to \cite{kondapally2022}, a lower boundary for the completeness corrections at a factor 10. This threshold ensures that the corrections do not become unreliable large for sources at the lower flux density end of our catalogues. 
	
	In Appendix \ref{app:constructiontests} we discuss a few tests to validate the RLF construction method and test the reliability of our $V_{\text{max}}$ derivations and completeness corrections with a local sample. These tests show that both the $V_{\text{max}}$ and the completeness corrections are applied well and recover a good estimate of the space densities of FRIs and FRIIs.
	
	\section{Results}\label{sec:results}
	
	Gathering all the ingredients from Sections \ref{sec:redshiftsimulation} and \ref{sec:rlf}, we construct RLFs to study the FRI and FRII space density evolution between $10^{24.5} \lesssim L_{150} \lesssim 10^{28.5}$~W~Hz$^{-1}$ out to $z=2.5$. In this section we will present our main results, where we will look separately at the local RLF and the RLF up to $z=2.5$.
	
	\subsection{Local radio luminosity function}\label{sec:lrlf}
	
\begin{figure*}[!h]
		\centering
		\includegraphics[width=0.7\linewidth]{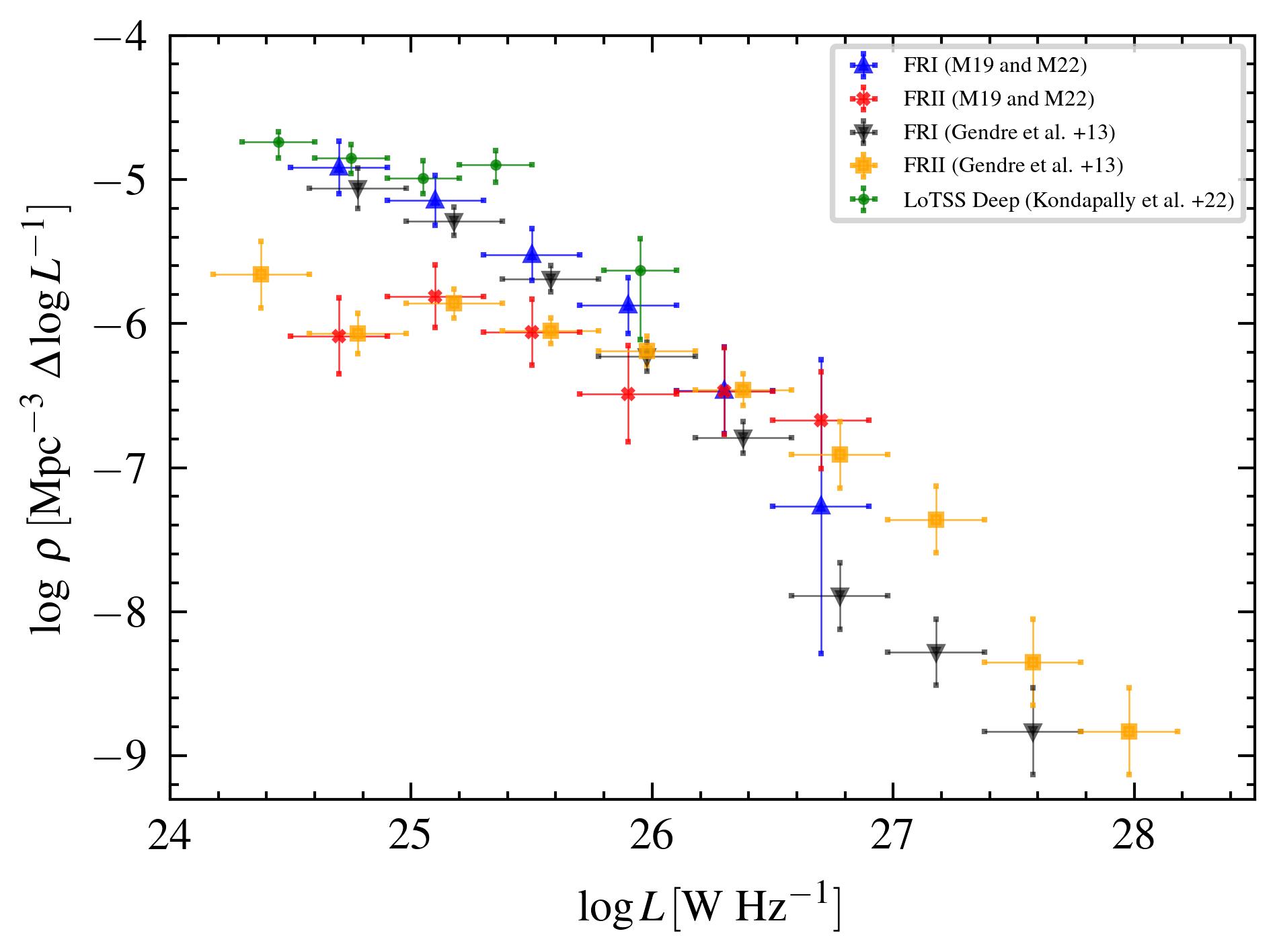}
		\caption{The local RLF for FRIs and FRIIs from this paper are plotted in blue and red respectively. For comparison and to validate that our FRI and FRII space densities do not extend above the total RLAGN population, we added the total local RLAGN RLF from \cite{kondapally2022} beyond $L_{150} \sim 10^{24}$~W~Hz$^{-1}$ in green. This total RLAGN RLF is constructed based on sources from the LoTSS-Deep Fields DR1. To demonstrate that our results agree with previous work, we also added the FRI and FRII RLF from \cite{gendre2013} in black and orange respectively. These points are based on the CoNFIG catalogue \citep{gendre2008, gendre2010}, where we converted the luminosity bins to 150 MHz by using a spectral index of $\alpha=0.7$. The $1\sigma$ bars from our RLFs are including $z_{\text{max}}$ errors, Poisson errors, completeness corrections errors, and classification errors. All our data points are given in Table \ref{table:localrlf}.}
		\label{fig:localrlf}
	\end{figure*}

	The local RLF constructed with the M19 and M22 sources below $z=0.3$ is shown in Figure \ref{fig:localrlf}. In this same figure we also plotted the FRI and FRII RLFs from \cite{gendre2013} and the total `radio-excess AGN' RLF from \citet[][hereafter referred to as the total RLAGN RLF]{kondapally2022}. We observe how all our local FRI and FRII space densities are near or below the total RLAGN space densities, which is constructed based on all RLAGN form the LoTSS-Deep Fields DR1 \citep{kondapally2021, duncan2021, tasse2021, sabater2021}. This is an important requirement, as the FRIs and FRIIs are a subset of the total RLAGN population. The FRI and FRII RLFs from \citeauthor{gendre2013} also agree well with our RLFs, which is another indication that our completeness corrections are working well (see also the tests in Appendix \ref{app:constructiontests}). The FRIs dominate below the break luminosity ($L_{150}\sim 10^{26}$~W~Hz$^{-1}$), while FRIIs dominate above. This is consistent with prior research \citep{fr, parma1996, ledlow, gendre2010}. Similar to \cite{gendre2010, gendre2013}, we find a smooth transition from the FRI to FRII dominance around the original break luminosity, where the FRI RLF drops off more strongly towards the higher radio luminosities compared to the FRII RLF. This smooth transition is also consistent with the FRI and FRII populations observed by FIRST \citep{capetti2017a, capetti2017b}.
	
	We can quantify the steepness of the space density drop-off after the break luminosity by fitting a broken power-law. To constrain better the power-law, we include the FRI and FRII data points from \citeauthor{gendre2010} for the radio luminosities where our RLFs do not cover (see Figure \ref{fig:localrlf}). This broken power-law is given by
	\begin{equation}\label{powerlaw}
		\rho(L_{\nu}) = \rho_{0}\left(\left(\frac{L_{\nu}}{\Tilde{L}}\right)^{\alpha} + \left(\frac{L_{\nu}}{\Tilde{L}}\right)^{\beta}\right)^{-1},
	\end{equation}
	where $\alpha$ and $\beta$ are respectively the low and high luminosity exponents, $\rho_{0}$ the characteristic space density, and $\Tilde{L}$ is the break luminosity. In Table \ref{table:fit} we give our best fit parameters for both the local FRI and FRII fit. We see from these values how the local FRI space densities are declining more strongly towards higher radio luminosities compared to the FRIIs. We also find different values for $\Tilde{L}$, which is usually taken to be equal to the break luminosity and determined by visual inspection, as was done by \cite{gendre2010}. Despite this different choice, the values for $\Tilde{L}$ are still consistent with the break luminosity around \textasciitilde $L_{150} \sim 10^{26}$~W~Hz$^{-1}$ we find in the literature \citep[e.g.][]{fr, jackson1999, willott2001, kaiser2007, mingo2019}.

	\begin{table*}[h!]
		\begin{center}
			\begin{tabular}{ c | c c c c }
				& $\rho_{0}$ & $\alpha$ & $\beta$ & $\Tilde{L}$ \\ [0.5ex] 
				\arrayrulecolor{gray}\hline
				\textbf{FRI} & -5.5 Mpc$^{-3} \Delta\log{L^{-1}}$ & 0.59 & 2.0 & 25.8~W~Hz$^{-1}$ \\ 
				\textbf{FRII} & -6.15 Mpc$^{-3} \Delta\log{L^{-1}}$& 0.13 & 1.7 & 26.4~W~Hz$^{-1}$\\
				\hline
			\end{tabular}
			\caption{Best fit parameters for the local FRI and FRII fit from Equation \ref{powerlaw}.}
			\label{table:fit}
		\end{center}
	\end{table*}
	
	\begin{table*}[h!]
		\begin{adjustwidth}{-1.07cm}{}
			\begin{center}
				\begin{tabular}{|c| c c | c c | c c | c c | c c |}
					\arrayrulecolor{gray}\hline
					$\log{L}$ & \multicolumn{2}{c}{$0<z<0.3$} & \multicolumn{2}{c}{$0.3<z<0.5$} & \multicolumn{2}{c}{$0.5<z<0.8$} & \multicolumn{2}{c}{$0.8<z<1.5$} & \multicolumn{2}{c}{$1.5<z<2.5$} \\
					(W~Hz$^{-1}$) &        \textbf{FRI} &         \textbf{FRII} &         \textbf{FRI} &         \textbf{FRII} &         \textbf{FRI} &         \textbf{FRII} &         \textbf{FRI} &         \textbf{FRII} &           \textbf{FRI} &           \textbf{FRII} \\[0.15cm] \hline \hline
					24.7 &  $-4.92_{-0.33}^{+0.18}$ &  $-6.09_{-0.39}^{+0.26}$ &  $-5.26_{-0.42}^{+0.24}$ &  $-6.05_{-0.41}^{+0.24}$ &   $-6.32_{-0.5}^{+0.32}$ &  $-6.69_{-0.54}^{+0.38}$ &                        - &                        - &                        - &                        - \\
					&                    (114) &                      (9) &                     (94) &                     (21) &                     (13) &                      (7) &                         &                         &                         &                         \\
					25.1 &  $-5.15_{-0.32}^{+0.17}$ &  $-5.81_{-0.34}^{+0.22}$ &   $-5.28_{-0.36}^{+0.2}$ &  $-6.23_{-0.37}^{+0.23}$ &  $-5.64_{-0.45}^{+0.26}$ &  $-6.43_{-0.41}^{+0.25}$ &  $-6.37_{-0.74}^{+0.64}$ &  $-6.72_{-0.73}^{+0.64}$ &                        - &                        - \\
					&                     (86) &                     (16) &                    (146) &                     (18) &                     (84) &                     (24) &                      (2) &                      (2) &                         &                         \\
					25.5 &   $-5.52_{-0.3}^{+0.18}$ &  $-6.06_{-0.33}^{+0.23}$ &   $-5.4_{-0.34}^{+0.19}$ &  $-6.27_{-0.36}^{+0.22}$ &  $-5.73_{-0.38}^{+0.21}$ &  $-6.51_{-0.37}^{+0.22}$ &  $-6.17_{-0.44}^{+0.29}$ &  $-6.27_{-0.46}^{+0.31}$ &                        - &                        - \\
					&                     (40) &                     (12) &                    (130) &                     (18) &                    (133) &                     (26) &                     (10) &                      (8) &                         &                         \\
					25.9 &  $-5.87_{-0.29}^{+0.19}$ &   $-6.49_{-0.4}^{+0.33}$ &  $-5.88_{-0.33}^{+0.19}$ &  $-6.29_{-0.34}^{+0.22}$ &   $-5.9_{-0.34}^{+0.19}$ &  $-6.43_{-0.35}^{+0.21}$ &  $-6.33_{-0.41}^{+0.28}$ &   $-6.17_{-0.4}^{+0.25}$ &   $-6.8_{-0.74}^{+0.64}$ &  $-6.98_{-0.73}^{+0.64}$ \\
					&                     (21) &                      (5) &                     (46) &                     (19) &                    (119) &                     (35) &                      (9) &                     (13) &                      (2) &                      (2) \\
					26.3 &   $-6.46_{-0.36}^{+0.3}$ &   $-6.47_{-0.35}^{+0.3}$ &   $-6.6_{-0.33}^{+0.24}$ &  $-6.49_{-0.32}^{+0.22}$ &  $-6.36_{-0.33}^{+0.19}$ &   $-6.49_{-0.33}^{+0.2}$ &  $-7.06_{-0.67}^{+0.61}$ &  $-6.23_{-0.37}^{+0.25}$ &  $-7.04_{-0.57}^{+0.47}$ &   $-7.2_{-0.69}^{+0.62}$ \\
					&                      (6) &                      (6) &                     (11) &                     (14) &                     (46) &                     (34) &                      (2) &                     (13) &                      (3) &                      (2) \\
					26.7 &  $-7.27_{-1.03}^{+1.02}$ &  $-6.67_{-0.38}^{+0.34}$ &   $-6.91_{-0.36}^{+0.3}$ &   $-6.61_{-0.3}^{+0.23}$ &  $-6.97_{-0.33}^{+0.23}$ &  $-7.02_{-0.33}^{+0.23}$ &  $-7.37_{-1.06}^{+1.03}$ &  $-6.78_{-0.48}^{+0.39}$ &  $-7.26_{-0.68}^{+0.61}$ &  $-6.82_{-0.47}^{+0.36}$ \\
					&                      (1) &                      (5) &                      (6) &                     (12) &                     (13) &                     (12) &                      (1) &                      (4) &                      (2) &                      (5) \\
					27.1 &                        - &                        - &  $-7.73_{-1.04}^{+1.02}$ &  $-7.03_{-0.38}^{+0.33}$ &                        - &   $-6.8_{-0.29}^{+0.19}$ &  $-7.44_{-1.05}^{+1.02}$ &  $-6.73_{-0.43}^{+0.34}$ &  $-7.63_{-1.06}^{+1.03}$ &  $-6.58_{-0.38}^{+0.26}$ \\
					&                         &                         &                      (1) &                      (5) &                         &                     (22) &                      (1) &                      (5) &                      (1) &                     (11) \\
					27.5 &                        - &                        - &                        - &  $-7.75_{-1.03}^{+1.02}$ &   $-7.88_{-0.63}^{+0.6}$ &   $-7.42_{-0.35}^{+0.3}$ &                        - &  $-7.51_{-1.04}^{+1.02}$ &                        - &    $-6.81_{-0.4}^{+0.3}$ \\
					&                         &                         &                         &                      (1) &                      (2) &                      (6) &                         &                      (1) &                         &                      (7) \\
					27.9 &                        - &                        - &                        - &  $-7.76_{-1.03}^{+1.02}$ &                        - &   $-7.93_{-0.62}^{+0.6}$ &                        - &  $-7.06_{-0.49}^{+0.45}$ &                        - &   $-7.41_{-0.65}^{+0.6}$ \\
					&                        &                        &                         &                      (1) &                         &                      (2) &                         &                      (3) &                         &                      (2) \\
					28.3 &                        - &                        - &                        - &  $-7.76_{-1.03}^{+1.02}$ &                        - &                        - &                        - &                        - &                        - &  $-7.78_{-1.04}^{+1.02}$ \\
					&                         &                         &                         &                      (1) &                         &                         &                         &                         &                        &                      (1) \\[0.15cm]
					\hline
				\end{tabular}
				\caption{RLF space density values in Mpc$^{-3} \Delta\log{L^{-1}}$ from Figures \ref{fig:localrlf} and \ref{fig:rlf_totalagn}. The source count is given between brackets below each space density value.}
				\label{table:localrlf}
			\end{center}
		\end{adjustwidth}
	\end{table*}
	
	\subsection{Radio luminosity function up to $z=2.5$}\label{sec:deeprlf}
	
\begin{figure*}[!h]
		\centering
		\includegraphics[width=0.9\linewidth]{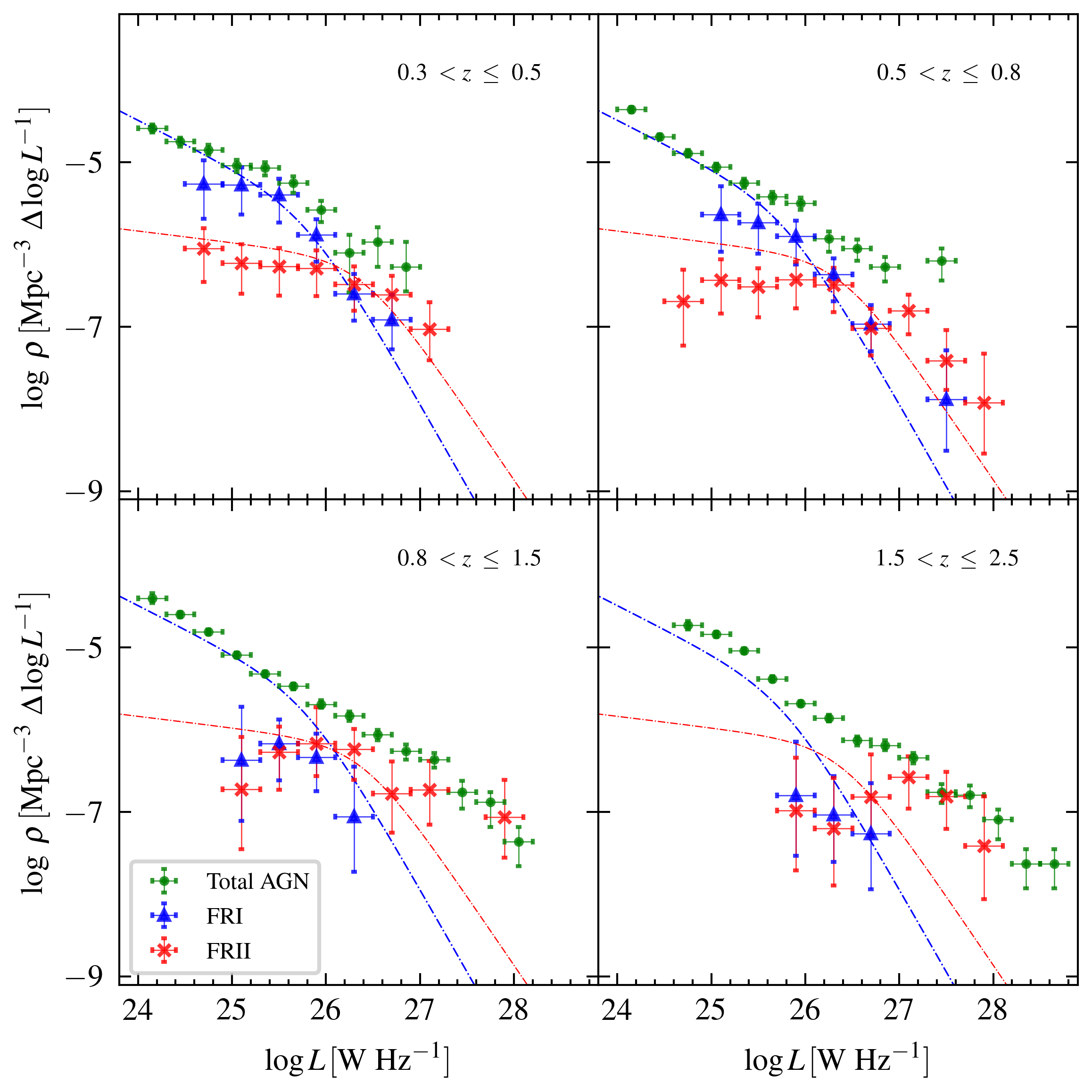}
		\caption{RLF for FRI (blue) and FRII (red) sources split in different redshift bins. For comparison and to show that our FRI and FRII space densities do not extend above the total RLAGN population, we added the total RLAGN RLF from \cite{kondapally2022} beyond $L_{150} \sim 10^{24}$~W~Hz$^{-1}$ in green. We also added the local RLFs for both FRIs and FRIIs with dot-dashed lines. We removed radio luminosity bins which only contain a single source.
			The $1\sigma$ bars from our RLFs are including $z_{\text{max}}$ errors, Poisson errors, completeness corrections errors, and classification errors, as explained in Sections \ref{sec:vmax} and \ref{sec:completeness}. Data points are given in Table \ref{table:localrlf}.}
		\label{fig:rlf_totalagn}
	\end{figure*}

\begin{figure*}[!h]
		\centering
		\includegraphics[width=0.48\linewidth]{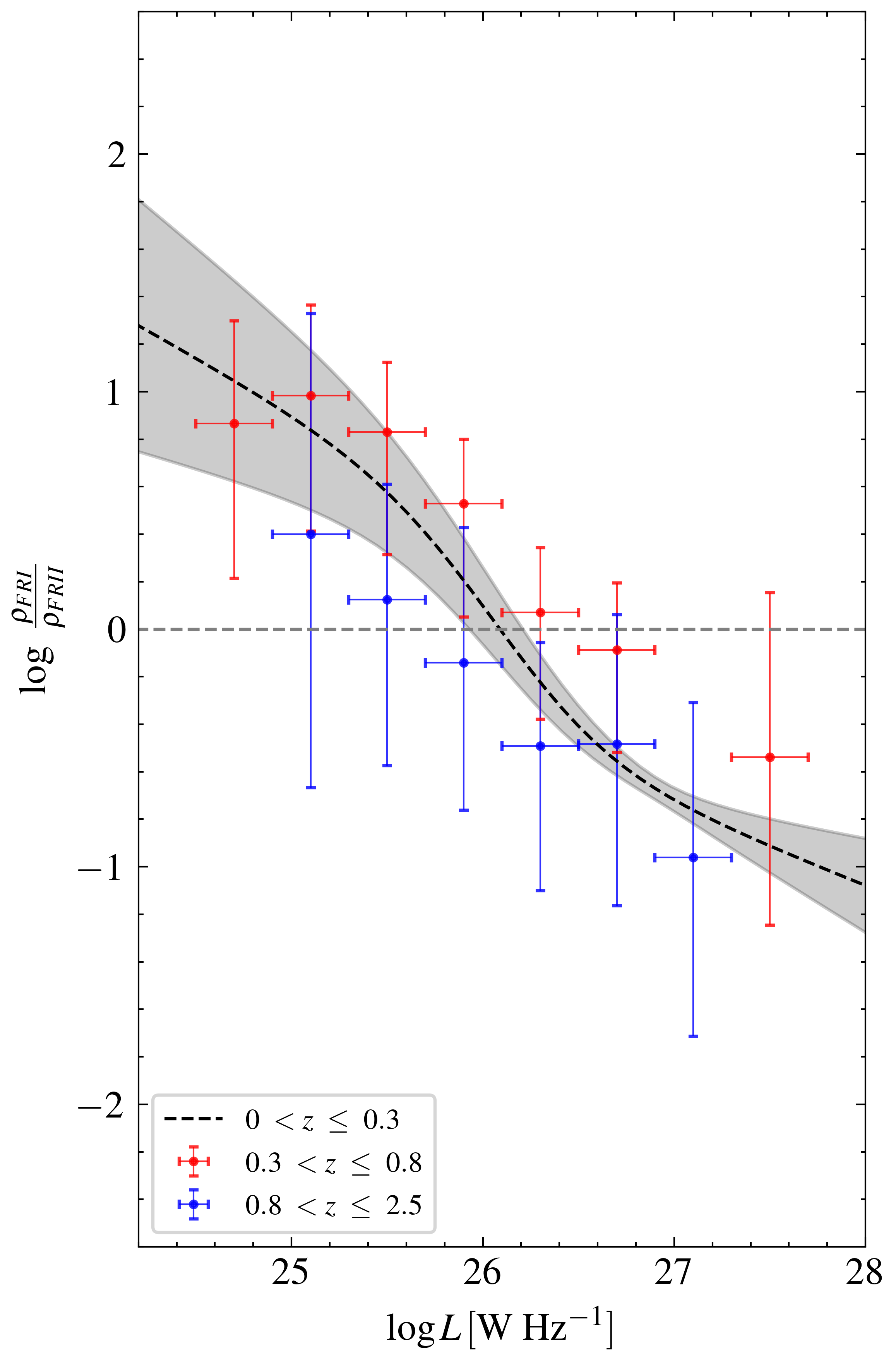}
		\centering
		\includegraphics[width=0.48\linewidth]{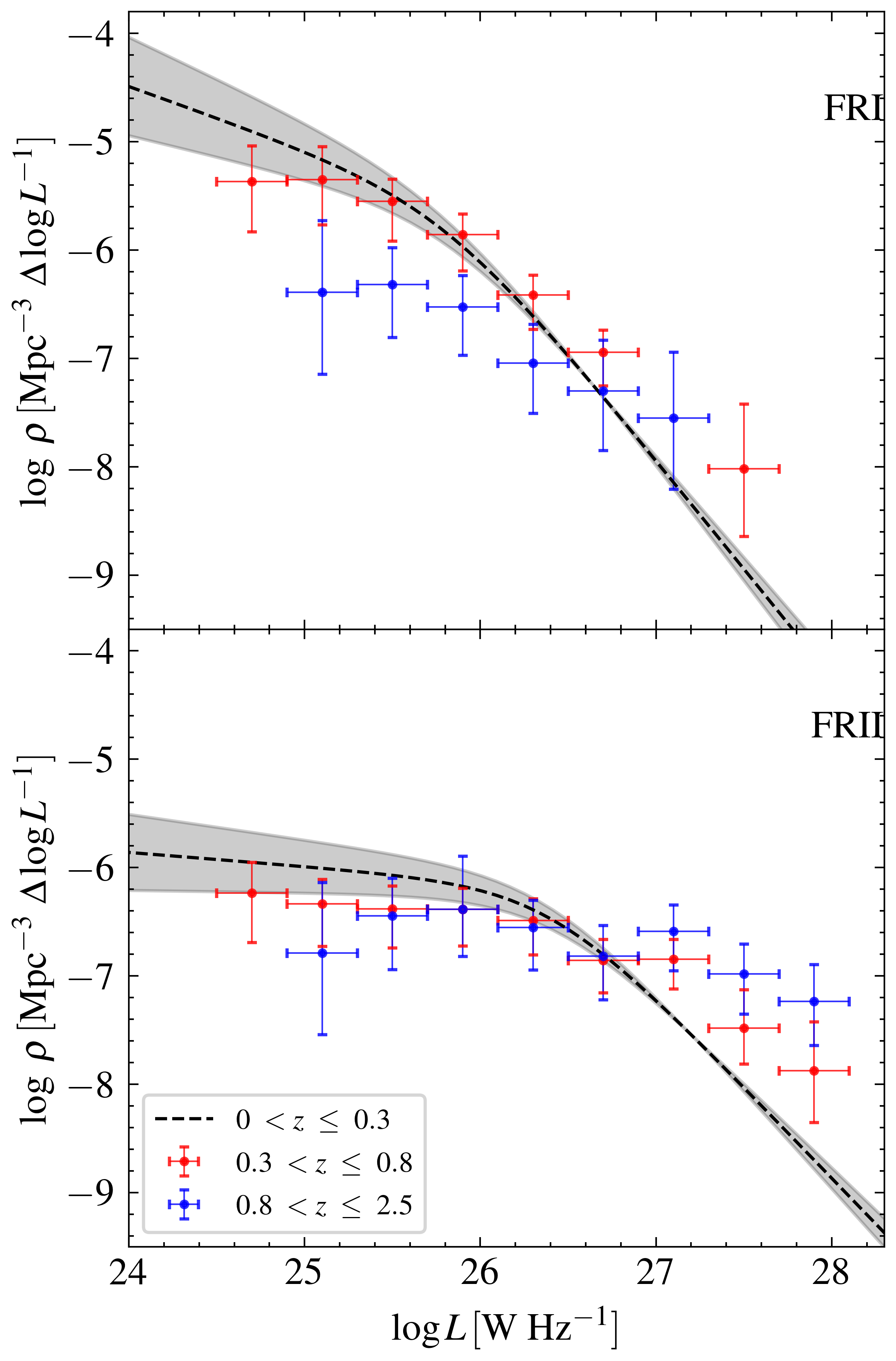}
		\caption{RLF ratio evolution. \textit{Left panel:} Ratio evolution over different redshift bins. \textit{Right panel:} Separated FRI (above) and FRII (below) space density evolution plots corresponding to the ratio evolution plot in the left panel. In both figures we added the \cite{gendre2013} data points for the higher end of the luminosity range, such that we can further constrain the local RLF, which was fitted with a broken power law from Equation \ref{powerlaw} and corresponding values from Table \ref{table:fit}, where we used a $1\sigma$ confidence interval. We removed radio luminosity bins which only contain a single source.
			The $1\sigma$ bars from our RLFs are including $z_{\text{max}}$ errors, Poisson errors, completeness corrections errors, and classification errors, as explained in Sections \ref{sec:vmax} and \ref{sec:completeness}.}
		\label{fig:rlfratio}
	\end{figure*}

	In Figure \ref{fig:rlf_totalagn} we see both the FRI and FRII space density as well as the total RLAGN from \cite{kondapally2022}. 
	Similar to the local RLF in Figure \ref{fig:localrlf}, we find in Figure \ref{fig:rlf_totalagn} that the space densities from the FRI and FRII samples are below the total RLAGN RLF. We see for the redshift bins below $z=0.8$ a break luminosity, where FRIs dominate below and FRIIs above $L_{150}\sim 10^{26}$~W~Hz$^{-1}$. Beyond $z=0.8$ the break luminosity becomes less well-defined. 
	In addition to Figure \ref{fig:rlf_totalagn}, we compare in the right panel from Figure \ref{fig:rlfratio} the space density evolutions per morphology class for 3 redshift bins, where the local RLF is fitted with a broken power-law as explained in the previous subsection. By subtracting the space densities from the FRIs and FRIIs in log space and propagating their error bars, we get the FRI/FRII ratio evolution in the left panel of this same figure. We combined all sources between $z=0.8$ and $z=2.5$ in one redshift bin because we have only 112 out of the total 1560 sources in this redshift range.
	
	We find in Figure \ref{fig:rlfratio} the FRII space densities to decrease at a lower rate as a function of radio luminosity when we compare the highest with the lowest redshift bin.
	This results above $L_{150}\sim 10^{27}$~W~Hz$^{-1}$ in a clear space density enhancement across redshift by $\sim 0.6-1\,\rm{dex}$ between the local and $z>0.8$ RLFs.
	Also \cite{gendre2010} found for FRIIs a similar space density enhancement above $L_{150} \sim 10^{27}$~W~Hz$^{-1}$, where we converted their luminosities from 1.4~GHz to 150~MHz with a spectral index of $\alpha=0.7$. They also report space density enhancements for FRIs at those radio luminosities, which also agrees with the mild enhancements that we detect.
	
	In the right panel of Figure \ref{fig:rlfratio} we detect a space density decline for FRIs from the local ($z<0.3$) to the highest redshift bin ($0.8<z<2.5$) below $L_{150} \sim 10^{26}$~W~Hz$^{-1}$. For FRIIs we do not find a significant space density decline below this radio luminosity. The space density decline for FRIs is not reported by \cite{gendre2010}.
 Nonetheless, both our results do within errors still agree with each other, as \citeauthor{gendre2010} only relied on 7 FRIs with associated redshifts above $z=0.3$, making their uncertainties large.
	Despite the FRI space density decline beyond $z=0.8$, we find the FRI/FRII space density ratios as a function of radio luminosity to remain fairly constant within error bars in the left panel of Figure \ref{fig:rlfratio}.
	
\begin{figure*}[!h]
		\centering
		\includegraphics[width=0.7\linewidth]{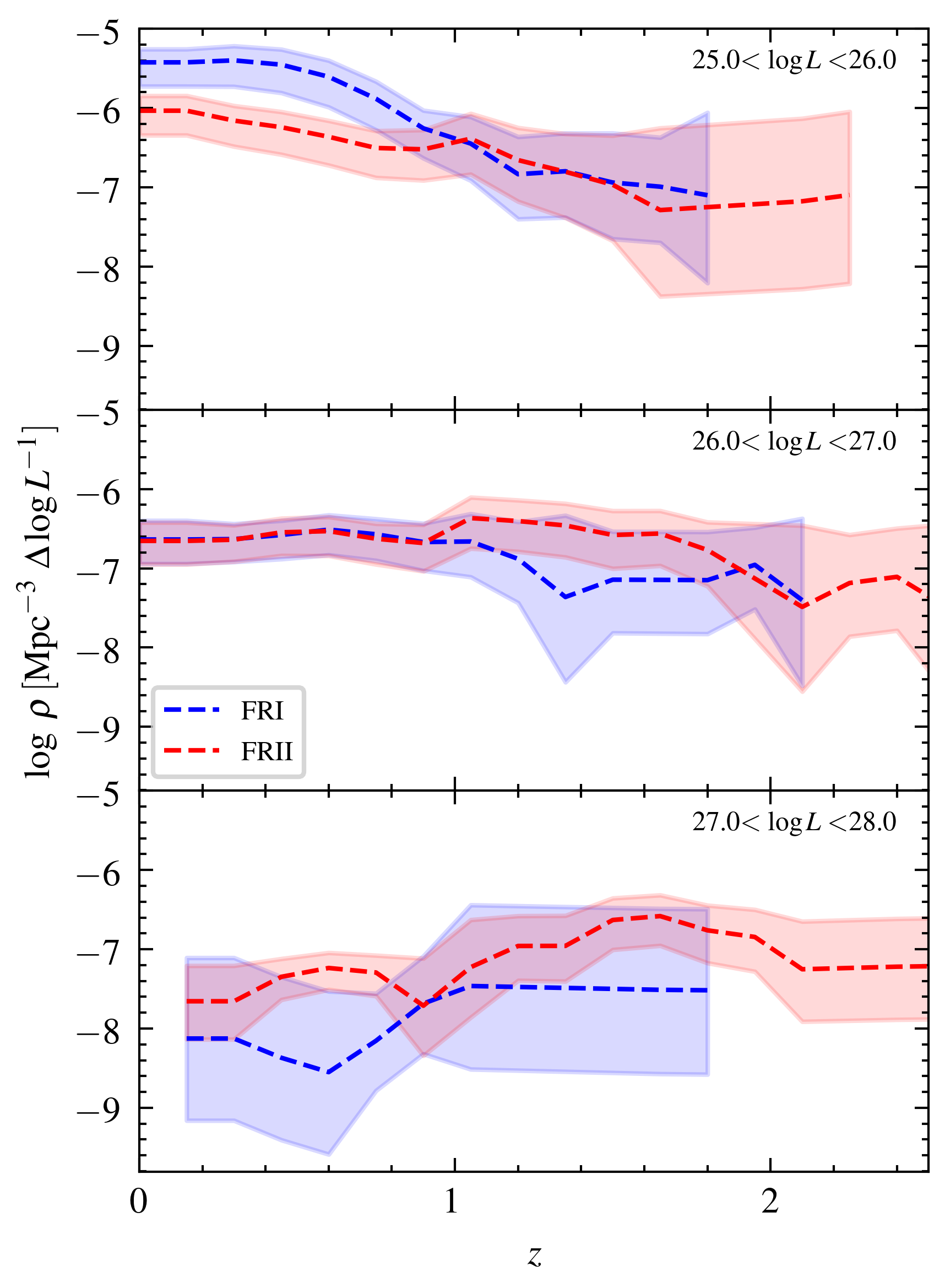}
		\caption{Space density evolution as a function of redshift ($z$) for 1 dex luminosity bins. This is constructed with sources from M19 and M22. The $1\sigma$ bars from our RLFs are including $z_{\text{max}}$ errors, Poisson errors, completeness corrections errors, and classification errors, as explained in Sections \ref{sec:vmax} and \ref{sec:completeness}.}
		\label{fig:zevolution}
	\end{figure*}

	The different space density evolutions as a function of redshift are also shown in Figure \ref{fig:zevolution}, where we inverted the RLFs to space density evolutions as a function of redshift for radio luminosity bins.
	Between $10^{27}\lesssim L_{150} \lesssim 10^{28}$~W~Hz$^{-1}$ we find a (mild) space density enhancement for both FRIs and FRIIs, which finds its maximum around $z=1.6$. The space densities for FRIs and FRIIs do not show any notable redshift evolution between $10^{26} \lesssim L_{150} \lesssim 10^{27}$~W~Hz$^{-1}$. For FRIs we find a mild space density decline above $z\sim 0.6$ and between $10^{25} \lesssim L_{150} \lesssim 10^{26}$~W~Hz$^{-1}$, while FRIIs do show for these radio luminosities hints of a mild declining trend above $z\sim 1$. However, this declining FRII space density trend is within error bars less significant than the declining trend of FRIs.
	
	\section{Discussion}\label{sec:discussion}
	
	With the arrival of larger FRI/FRII catalogues sensitive to lower flux densities and better redshift measurements, we have in the previous sections been able to construct RLFs up to $z=2.5$ based on the M19 and M22 catalogues.
	We will in this section discuss the effects of energy losses on our RLFs, how the total RLAGN RLF compares to our FRI and FRII RLFs, remaining selection biases, and how we can interpret the space density evolution that we find for FRIs and FRIIs.
	
	\subsection{Energy losses and the RLF}\label{sec:interpretingrlfs}
	
	At higher redshifts, IC losses for each source become more important, as more electrons will be scattered by CMB photons when the energy density of the CMB increases. These IC losses are suggested to be one of the primary mechanisms behind the $\alpha-z$ relation \citep{klamer2006, ghisellini2014, morabito2018, sweijen2022b}, which associates steeper redshifts with higher spectral indices \citep{tielens1979, debreuck2000, miley2008}. Because we are interested in the redshift evolution of FRIs and FRIIs, it is important to discuss how these energy losses affect the interpretation of our RLFs.
	
	The impact of the IC losses on the measured surface brightness of a source is mainly contingent upon the magnetic field strength within the lobes and the equivalent magnetic field associated with the CMB \citep{harwood2013}. 
	Due to their different morphologies, we anticipate the IC losses to affect FRIs and FRIIs disproportionately, as this effect can alter both the size and detectability of FRIs, whereas for FRIIs the detectability will likely not change because compact hot spots at the ends of the jets are relatively unaffected. 
	As a result, we expect that when we would take IC losses into account, the $z_{\text{max}}$ and the completeness corrections will change more significant for FRIs than for FRIIs. Similarly, we could also consider other losses, such as the effects of synchrotron losses on the measured radio flux densities of FRIs and FRIIs \citep{myers1985, alexander1987, jamrozy2008, harwood2013, sweijen2022b}. In addition, the more distant universe is also denser, which leads to more confinement of the radio lobes and lower adiabatic losses of sources at higher redshifts compared to sources in the local universe \citep[e.g.][]{barthel1996}. Nevertheless, incorporating corrections for all the above mentioned energy losses would violate our initial assumption that the radio luminosity from a source remains constant as a function of redshift while calculating $V_{\text{max}}$ (see Section \ref{sec:redshifting}). Therefore, for the interpretation of the comparison of FRIs and FRIIs in our RLFs it is important to be aware that these are measured radio luminosity functions and not energy loss-corrected radio luminosity functions.
	
	Fortunately, the FRI/FRII ratio RLF is less affected by those losses, as part of their effect on the $z_{\text{max}}$ and completeness corrections cancel out. As a result, this makes the physical interpretation more comparable to the hypothetical energy-loss adjusted RLFs. The left panel of Figure \ref{fig:rlfratio} supports this comparability, as the three redshift bins are the same within uncertainties.

	\subsection{Comparing FRI/FRII RLFs}\label{sec:totalagn}
	
	It is well known that FRIs and FRIIs make up a significant fraction of the total RLAGN population \citep{sadler2016, baldi2018}. In this subsection, we will discuss the comparison of our FRI and FRII RLFs with those from \cite{gendre2010, gendre2013} and \cite{kondapally2022}.
	For the total RLF we exclusively use the total RLAGN RLF from \citeauthor{kondapally2022}, as they demonstrated consistency with previous total RLAGN RLFs in the literature.
	
	Figure \ref{fig:localrlf} shows how local the FRI space densities closely follow the total RLAGN space densities up to about $L_{150} \sim 10^{26}$~W~Hz$^{-1}$, after which FRIIs become the dominant morphology population. This similarity between the FRI and total RLAGN space densities is due to the fact that the fraction of compact sources increases towards the lower radio luminosities and we only consider sources above $L_{150} \sim 10^{24.5}$~W~Hz$^{-1}$ \citep[e.g.][]{baldi2015, baldi2018, capetti2020}. It is reassuring to find the local RLF from \cite{gendre2013} to be similar to ours, as this indicates that the local RLFs constructed in this paper and by \citeauthor{gendre2013} are not biased by the data or RLF construction methods. In Figure \ref{fig:rlf_totalagn}, we observe that the space densities of the FRI and total RLAGN space densities continue to show consistent similarity up to a similar break luminosity and $z=0.8$. Examination of the FRI evolution above $z=0.3$ could not be accurately done by \cite{gendre2010}, due to the fact they had in their sample only 7 FRIs associated with redshifts above $z=0.3$ \citep[see also Figure 6 in][]{gendre2010}. This is over 100 times less FRIs compared to what we have available from the combined M19 and M22 catalogues, comprising 877 FRIs above $z=0.3$ with accurate redshifts \citep{duncan2019, duncan2021}.
	
	Beyond $z=0.8$ we do not find a well-defined break luminosity above which FRIIs dominate over FRIs. However, we do find in Figure \ref{fig:rlf_totalagn} above $L_{150} \sim 10^{26}$~W~Hz$^{-1}$ the FRIIs to be a dominant morphology type when we compare the total RLAGN space densities with the FRII space densities. This demonstrates that up to $z=2.5$ the FRIIs dominate the RLAGN population at the high radio luminosity end. Below $L_{150} \sim 10^{26}$~W~Hz$^{-1}$ and beyond $z=0.8$ we detect a prominent offset in the order of 1 to 1.5 dex between the total RLAGN and the FRI and FRII space densities. This is likely related to remaining unaccounted selection effects. Some of this might be a result of energy loss effects that we did not take into account, which more strongly affect FRI detections (see Section \ref{sec:interpretingrlfs}). Also the probable underestimation of the completeness corrections for FRIs due to the difficulty to accurately measure their sizes in the higher redshift bins could play a role (see Section \ref{sec:angularsizecorrections}). Moreover, \cite{kondapally2022} discusses how the space densities of star-forming low-excitation radio galaxies (LERGs) increase over redshift and dominate the LERG space densities after $z\sim 1$, while the space densities of quiescent LERGs decreases over redshift. Given that low-power FRIs and FRIIs are predominantly associated with LERGs \citep{mingo2022}, there is a possibility of an unaccounted selection bias in our completeness corrections if some of those star-forming LERGs might actually be FRIs or FRIIs with small jets that are difficult to detect at a resolution of 6\arcsec. We also argue that the difficulty to detect star-forming radio galaxies with radio jets in our sample is likely a more significant issue for FRIs, as their smaller weaker jets are more difficult to detect compared to FRIIs. Other missing morphology types like hybrid or double-double sources are expected to be rare \citep{harwood2020, mingo2019, mingo2022} and will therefore not have a significant contribution to remaining selection effects. Thus, although the redshift simulations help us to correct for observational biases when deriving $z_{\text{max}}$ values (see Section \ref{sec:zmax}) and completeness corrections (see Section \ref{sec:cor1}), it is possible that residual resolution and surface brightness limits could still reside in selection effects that we have yet not been able to fully correct for.

	\subsection{Space density enhancements}\label{sec:spacedensityenh}
	
	Space density enhancements of the total RLAGN RLF have been well-measured for a long time \citep[e.g.][]{dunlop1990}. Multiple studies have found space density enhancements for bright FRIs and FRIIs \citep{snellen2001, willott2001, jamrozy2004, rigby2008, gendre2010}. Given our increased number of sources above $z=0.3$ compared to previous studies, it is valuable to re-examine the space density enhancements and declines with our RLFs and look specifically how these relate to our understanding of radio galaxy evolution.
	
	The space density enhancement from low to high redshift that we find for FRIs and FRIIs above $L_{150}\sim 10^{27}$~W~Hz$^{-1}$ could simply be related to the higher gas availability in the earlier universe, increasing the probability for RLAGN to become more powerful. Although taking into account energy loss effects in our RLFs is complex and out of the scope of this paper (see Section \ref{sec:interpretingrlfs}), it is worth mentioning that if the net energy loss effects are lower in the earlier denser universe (e.g. significant lower adiabatic losses) compared to the local universe, we could expect to measure even more powerful sources at higher redshifts. 
	We are not well-enough constrained for bright FRIs to find a notable difference between the space density enhancements of FRIs and FRIIs above $L_{150}\sim 10^{27}$~W~Hz$^{-1}$ (see Figures \ref{fig:rlfratio} and \ref{fig:zevolution}). Nevertheless, a recent simulation demonstrated how radio sources were more likely to live in less rich environments at $z=2$ compared to $z=0$ \citep{thomas2022}, which is in favor of FRII morphologies \citep{croston2019}. More data at the high radio luminosity end would help to better constrain the FRI and FRII space density enhancements and test if powerful FRIIs have indeed a stronger space density enhancement than powerful FRIs.
	
	The space density decline of FRIs below $L_{150}\sim 10^{26}$~W~Hz$^{-1}$ is likely related to unaccounted selection effects, as we discussed in Section \ref{sec:totalagn} to explain the observed space density offsets between the local and higher redshift bins. However, despite the fact that the remaining selection effects have a stronger effect on the space densities from FRIs than FRIIs, we find the FRI/FRII space density ratio over all radio luminosities to be remarkably similar when we compare the three redshift bins in the left panel of Figure \ref{fig:rlfratio}. If this holds for larger and more comprehensive sample sizes, resulting in reduced Poisson errors, it could suggest that the jet-disruption of FRIs is not primarily influenced by environmental factors, as those evolve over redshift. Instead, disruption of the jets from the FRI parent population would more likely be associated with events happening in close proximity to, or within, the parent radio galaxy.
	
	\subsection{Future prospects}\label{sec:future}
	
	We demonstrate in this paper how detection limits are affecting selection biases and how it is possible to partly correct for these if we want to construct reliable RLFs from RLAGN. In order to enhance the significance of our findings, we need more objects from more sensitive radio maps along with wider area observations. The 40\arcsec~angular size cut that we needed to apply, due to the angular size selection from \cite{mingo2019}, removes a large fraction of the small FRIs and FRIIs. Especially sources at the lowest radio flux densities are the most affected. The angular size cut is due to the 6\arcsec~resolution from the LoTSS radio maps that were used to extract the FRI and FRII sources from. So, to collect a reliable sample below 40\arcsec, we need to improve the resolution.
	
	Recent efforts have been ongoing to increase the sky coverage with LoTSS DR2 at 6\arcsec~\citep{shimwell2022}, which will increase the number of objects in M19 by a factor of about \textasciitilde 13. This data release is based on radio maps at the same resolution and a similar sensitivity as LoTSS DR1. This will therefore only help to reduce the uncertainties at the higher radio luminosity end and up to $z=0.8$. Fortunately, work has also been done to bring LOFAR's resolution down to 0.3\arcsec~by including all baselines up to \textasciitilde 2000~km \citep{varenius2015, varenius2016, harris2019, morabito2022, sweijen2022}. This resolution will help us to collect sources that are unresolved at 6\arcsec, such that we can lower the 40\arcsec~angular size cut.
 	An improved resolution comes with the unavoidable loss of surface brightness sensitivity. \cite{sweijen2022} reported from 6\arcsec to 0.3\arcsec an estimated 60\% detection loss of sources that were unresolved at 6\arcsec. As FRIs are on average less bright and have by definition diffuser jets than FRIIs, this will surely have a more negative effect on the detections of FRIs compared to FRIIs. To ensure that enough small FRIs will be detected, it becomes essential to improve the sensitivity by processing and calibrating multiple observations of the same fields (de Jong et al. in prep.) and to complement these high resolution radio maps with intermediate resolutions around for example \textasciitilde1\arcsec \citep{ye2023}.

	\section{Conclusions}\label{sec:conclusion}
	
	We presented in this paper RLFs of FRI and FRII morphologies up to $z=2.5$ and beyond $L_{150}\sim 10^{24.5}$~W~Hz$^{-1}$, by utilizing the M19 and M22 catalogues. We corrected for redshift effects and the incompleteness of our sample by using the \texttt{redshifting} algorithm. This algorithm gave us a reliable estimate of the maximum distance an FRI and FRII source can still be classified. In particular, our RLFs for FRIs are an improvement above $z=0.3$, as we have over 100 times more available FRIs with associated redshifts above this redshift, compared to previous studies \citep{gendre2010, gendre2013}. These RLFs also provide us with continuing evidence of evolution of FRI and FRII sources.
	
	In our RLFs we do not detect any sharp transitions between the FRI and FRII morphologies as a function of radio luminosity or redshift. We find a space density enhancement from low to high redshift for FRI and FRII sources beyond $L_{150}\sim 10^{27}$~W~Hz$^{-1}$, which might be explained by a higher gas availability in the earlier universe, increasing the likelihood for FRIs and FRIIs to become powerful. Also, the net energy losses at higher redshifts could potentially increase the measured radio luminosities at higher redshifts if the adiabatic losses are significantly lower compared to lower redshifts. However, to test this assertion we need a more detailed analysis of the different energy loss effects (e.g. IC and synchrotron losses), which is out of the scope of this paper. At the low radio luminosity end, we tentatively identify a declining trend of the FRI space densities with redshift. This is likely related to remaining selection effects such as from the underestimation of distant FRI angular sizes at higher redshifts and the difficulty to detect star-forming FRIs with small jets, which have been proven to be more prominent at higher redshifts. 	Although there are significant uncertainties represented by the large error bars, the evolution of the FRI/FRII ratio that we derive suggests that the FRI morphology is primarily a result of the disruption of jets on scales originating within or close to the host galaxy, rather than jet-disruption due to large-scale environmental factors.
	
	The potential residual selection biases in our results highlight the necessity to further develop the calibration and imaging pipeline of LOFAR data with baselines up to 2000~km, such that it will be possible to incorporate radio sources at smaller angular scales and lower radio luminosities. This will eventually help to more precisely model the FRI and FRII sources above $z=0.3$ towards lower radio luminosities, which can further enhance our understanding of radio jet evolution from RLAGN and the link between these jets and their environment and host galaxies.
	
	\begin{acknowledgements} 
		
		This publication is part of the project CORTEX (NWA.1160.18.316) of the research programme NWA-ORC which is (partly) financed by the Dutch Research Council (NWO). This work made use of the Dutch national e-infrastructure with the support of the SURF Cooperative using grant no. EINF-1287.
		
		RK and PNB are grateful for support from the UK STFC via grant ST/V000594/1.
		
		BM acknowledges support from the Science and Technology Facilities Council (STFC) under grants ST/T000295/1 and ST/X001164/1.
		
		RJvW acknowledges support from the ERC Starting Grant ClusterWeb 804208. 
		
		This work was supported by the Medical Research Council [MR/T042842/1].
		
		LOFAR data products were provided by the LOFAR Surveys Key Science project (LSKSP; \url{https://lofar-surveys.org/}) and were derived from observations with the International LOFAR Telescope (ILT). LOFAR \citep{haarlem2013} is the Low Frequency Array designed and constructed by ASTRON. It has observing, data processing, and data storage facilities in several countries, which are owned by various parties (each with their own funding sources), and which are collectively operated by the ILT foundation under a joint scientific policy. The efforts of the LSKSP have benefited from funding from the European Research Council, NOVA, NWO, CNRS-INSU, the SURF Co-operative, the UK Science and Technology Funding Council and the Jülich Supercomputing Centre.
		
		For the purpose of open access, the author has applied a Creative Commons Attribution (CC BY) licence to any Author Accepted Manuscript version.

	\end{acknowledgements} 
	
	\bibliographystyle{bst}
	
	\bibliography{bib}
	
	\appendix

	\section{Testing the RLF reliability}\label{app:constructiontests}
	
\begin{figure*}[ht]
		\includegraphics[width=0.48\linewidth]{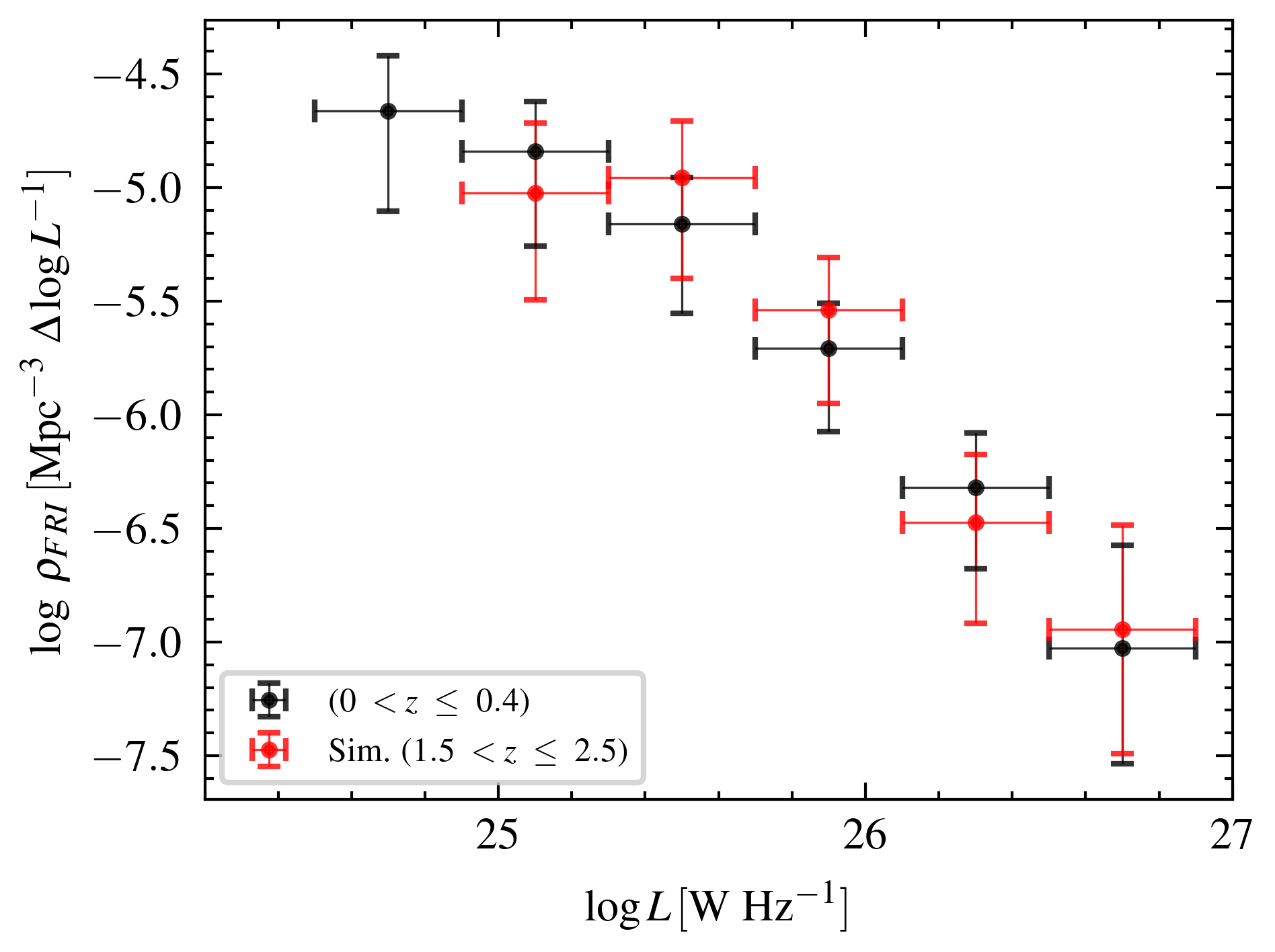}
		\includegraphics[width=0.48\linewidth]{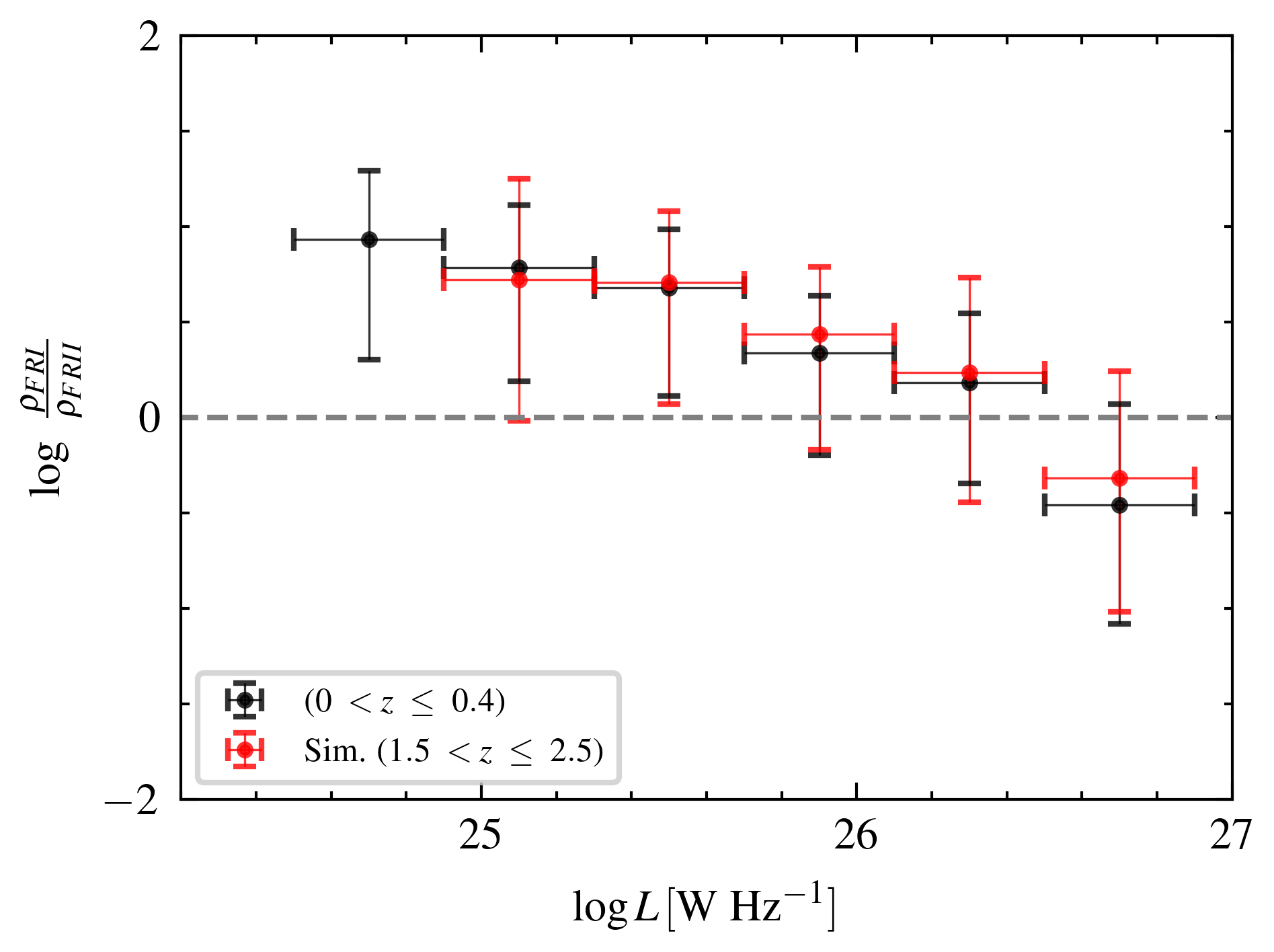}
		\caption{RLF space density evolution difference between the original and simulated sample. \textit{Left panel:} RLF for FRI sample before and after applying the \texttt{redshifting} algorithm from $z<0.4$ to $z=1.5$. \textit{Right panel:} RLF ratio for a sample of FRI and FRII sources before and after applying the \texttt{redshifting} algorithm from $z<0.4$ to $z=1.5$. 
			Error bars from our RLFs are including $z_{\text{max}}$ errors, Poisson errors, completeness corrections errors, and classification errors.}
		\label{fig:FRIFRIIsimulation}
	\end{figure*}

	Relocating sources to higher redshifts reduces their flux densities and apparent sizes, which makes it more difficult to detect and classify FRI and FRII sources. Figure \ref{fig:FRIFRII_redshift} shows how FRIs and FRIIs are affected differently by these effects. To test whether the $V_{\text{max}}$ method and our completeness corrections are taking the effect, due to changes in flux density and resolution, correctly into account, we compare an FRI and FRII sample below $z<0.4$ with the same sample shifted to $z=1.5$. This is done by using the algorithm from Section \ref{sec:redshifting} and applying the same completeness correction process discussed in Section \ref{sec:completeness}. If our $V_{\text{max}}$ method and completeness corrections are well-applied, we expect the derived space densities of the original $z<0.4$ sample and of the same sample shifted to $1.5<z<2.5$ to agree with each other, as the simulated sample has larger space density corrections and smaller $V_{\text{max}}$ values. The left panel from Figure \ref{fig:FRIFRIIsimulation} has a sample of 478 FRIs before relocating them to $z=1.5$. After relocating there are only 105 sources that could be classified. Although the number reduces with a factor 4, we find that the radio luminosity functions to still agree with each other. In the right panel of this same figure we also compare the RLF ratios before and after relocating and also find a strong agreement. We have also experimented with other redshift bins (smaller and larger) and find similar agreements. This gives us additional confidence that both the $V_{\text{max}}$ and the completeness corrections are applied well.
	
\begin{figure*}[!h]
		\centering
		\includegraphics[width=0.9\linewidth]{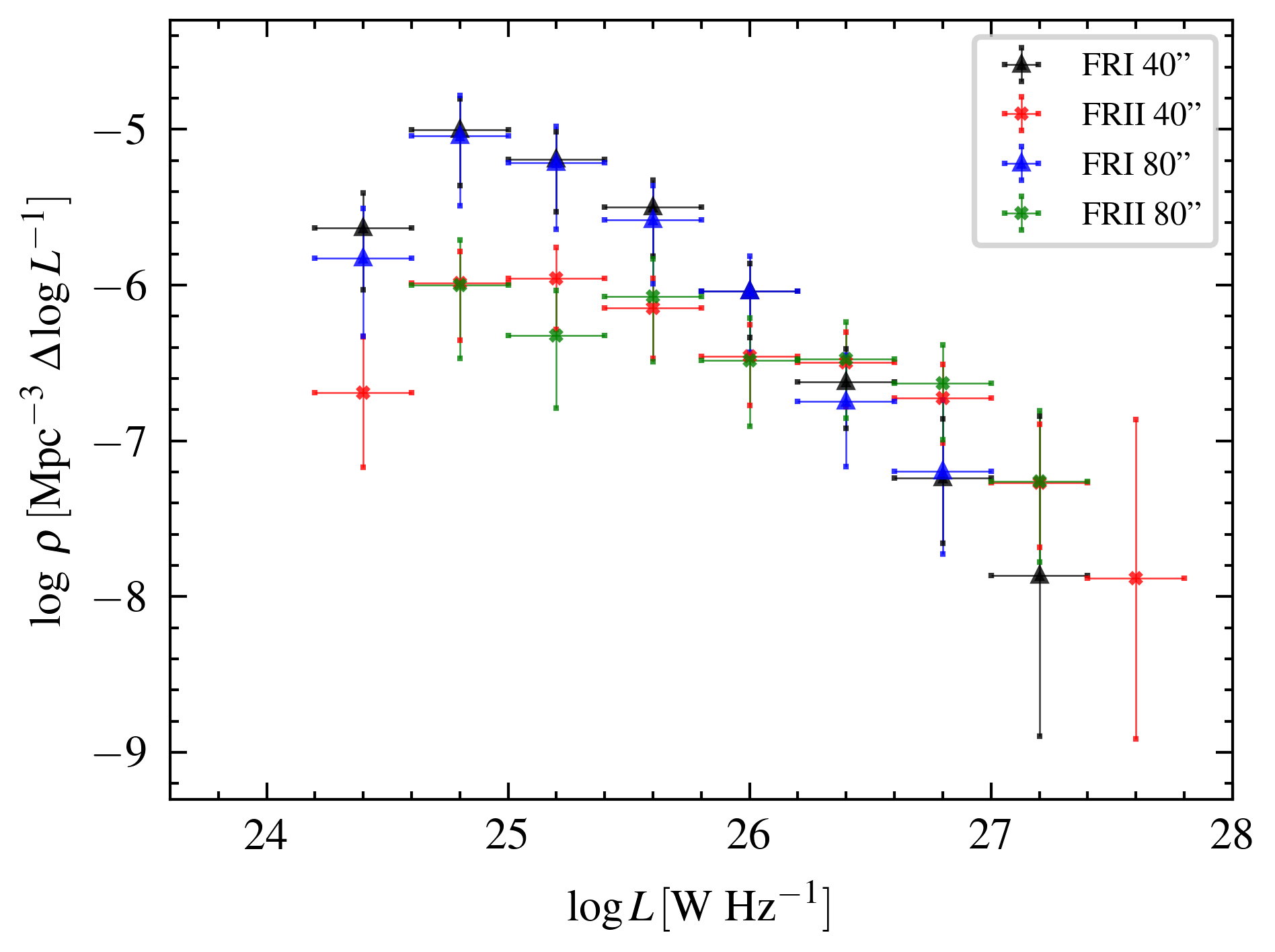}
		\caption{Comparison between a 40\arcsec~and 80\arcsec~source cut below $z=0.5$. Both RLFs are corrected with the completeness corrections from Section \ref{sec:completeness}. The angular size completeness correction is applied for respectively the 40\arcsec~and 80\arcsec~angular sizes. This figure contains all sources in the redshift bin $0<z<1.0$. The 40\arcsec cut selection has 865 sources, while the 80\arcsec~cut has 405 sources left. The 1$\sigma$ error bars are containing $z_{\text{max}}$ errors, Poisson errors, completeness corrections errors, and classification errors.}
		\label{fig:rlf_aseccut}
	\end{figure*}

	
	To test the completeness corrections derived with the angular size distribution (see Section \ref{sec:completeness}), we applied a larger angular size cut at 80\arcsec~and compared the space densities with the already present 40\arcsec~cut. The comparison for sources below $z=0.5$ is shown in Figure \ref{fig:rlf_aseccut} and shows how, although the number count reduces almost with a factor 2, the completeness corrections correct remarkably well for this effect and make the both RLFs agree with each other. We also compared this for other angular size cuts and found similar consistencies. This demonstrates the reliability to correct our space densities with the angular size distributions from \cite{windhorst1990} with updated fitting parameters from \cite{mandal2021} on a local sample.
	
\end{document}